\documentclass[twocolumn,secnumarabic,amssymb, nobibnotes, aps, pre, shortbibliography]{revtex4-2}

\newcommand{\teoadd}[1]{\textcolor{black}{#1}}

\newcommand{\JFadd}[1]{\textcolor{black}{#1}}

\usepackage{mathtools,amsmath}
\usepackage{tikz}
\usepackage{graphicx}
\usepackage{comment}
\usepackage{mdframed}
\usepackage{setspace}
\usepackage[mathlines]{lineno}
\newcommand{\f}[2]{\frac{#1}{#2}}
\newtheorem{theorem}{Theorem}

\makeatother
\begin{document}
%\linenumbers % line numbering

\title{Self\teoadd{-segregation} in heterogeneous \JFadd{metapopulation} landscapes}%
\author{Jean-François de Kemmeter$^1$, Timoteo Carletti$^1$, Malbor Asllani$^{2,3}$}
\affiliation{\vspace{.25cm}
$1$ naXys, Namur Institute for Complex Systems, \& Department of Mathematics, University of Namur, rue Graf\'e, 2 B5000, Belgium\\
$2$ School of Mathematics and Statistics, University College Dublin, Belfield, Dublin 4, Ireland\\
$3$ Department of Mathematics, Florida State University, 1017 Academic Way, Tallahassee, FL 32306, United States of America
}
\date{\today} %

%\doublespacing
\begin{abstract}
\noindent
Complex interactions are at the root of the population dynamics of many natural systems, particularly for being responsible for the allocation of species and individuals across apposite niches of the ecological landscapes. On the other side, the randomness that unavoidably characterises complex systems has increasingly challenged the niche paradigm providing alternative neutral theoretical models. We introduce a network-inspired \JFadd{metapopulation} individual-based model (IBM), hereby named \emph{self\teoadd{-segregation}}, where the density of individuals in the hosting patches (local habitats) drives the individuals spatial assembling while still constrained by nodes' saturation. In particular, we prove that the core-periphery structure of the networked landscape triggers the spontaneous emergence of vacant habitat patches, which segregate the population in multistable patterns of isolated \JFadd{(sub)communities} separated by empty patches. 
Furthermore, a quantisation effect in the number of vacant patches is observed once the total system mass varies continuously, emphasising thus a striking feature of the robustness of \JFadd{population} stationary distributions. \JFadd{Notably, our model reproduces the patch vacancy found in the fragmented habitat of the Glanville fritillary butterfly \emph{Melitaea cinxia}, an endemic species of the \r{A}land islands.}
We argue that such spontaneous breaking of \JFadd{the natural habitat} supports the concept of the highly contentious (Grinnellian) niche vacancy and also suggests a new mechanism for the endogeneous habitat fragmentation and consequently the peripatric speciation.   
\end{abstract}

\maketitle

\section{Introduction}

The distinct feature of complex systems is their critical reliance on the interactions between the constituent entities. Such strong dependence has given birth to the discipline of network science \cite{newman_2017_networks,boccaletti_2006_complex,porter_2016_dynamical,barrat2008dynamical}. From this perspective, nodes represent either single individuals or groups thereof and edges, the respective interactions among them. According to the positive, negative or neutral nature of such interactions, they can identify mutualism, competition, or commensalism, respectively \cite{murray_mathematical_2002, kauffman_1993}. There is a long tradition in population dynamics, further strengthened by complexity science, to attribute collective behaviors such as self-organisation or adaption, in biological or social systems, to interactions between entities (e.g., Lotka-Volterra models) \cite{murray_mathematical_2002, newman_2017_networks, kauffman_1993}.
In ecology, the coexistence of a large variety of species is thought to be possible because of their ability to occupy specific niches, namely the set of environmental factors (resources, prey/predators, competitors, abiotic components, etc.) matched with a given species, as a consequence of a long and highly selective evolution process \cite{macarthur_theory_2001, hubbell_unified_2001}.
	
An alternative approach adopted in recent years to tackle the question of species diversity is that of neutral theory, according to which organised patterns can occur exclusively due to the randomness of several factors that decisively condition species survivability \cite{caswell_community_1976, macarthur_theory_2001, hubbell_unified_2001,azaele_statistical_2016, alonso_merits_2006, black_stochastic_2012}. Stochastic processes are the common ground around which the modern neutral theory of biodiversity is erected \cite{caswell_community_1976, hubbell_unified_2001, azaele_statistical_2016}. In this regard, intra- and inter-specific interactions are considered simultaneously relevant, consisting in a striking difference of the neutral model from niche ones \cite{leibold2004metacommunity}. Thus elementary processes such as births/deaths or dispersal (emigration and immigration) result crucial in describing the collective dynamics emerging from  individual-based models (IBM). In the context of neutral theory, there have been several attempts to better understand the segregation of species in ecological niches in terms of the randomness of the decisive factors in action during the evolutionary process \cite{caswell_community_1976, kauffman_1993}. For instance, there is a consolidated debate whether the species fully occupy all the niches to overcome Gause's law of competitive exclusion \cite{murray_mathematical_2002, hubbell_unified_2001, azaele_statistical_2016} or either vacant niches can exist due to an optimal fitness landscape \cite{kauffman_1993}. According to the latter thesis, the fitness function that describes the adaption of the species to the niche domain, is rather rugged and multipeaked. This way, the randomness associated with the initial conditions will allow the individuals assembling around some (local) optima and eventually allow the emergence of vacant patches around the landscape minima. On the other side, ecological landscapes found in nature are highly fragmented and are often conditioned by different geographic, abiotic, or human factors \cite{hubbell_unified_2001, macarthur_theory_2001,nicholson2020,castorani2015connectivity,hanski1994metapopulation,orsini2008genetic,hanski2011eco,hanski2017ecological,opedal2020host}. Understanding the mechanisms of fragmentation of the ecological landscapes can give valuable insight into the the evolution of animals, plants, and other living beings in their respective habitats \cite{weir_ice_2004, dias_natural_2013, thompson_speciation_2018}. By anticipating on the following let us emphasize that the proposed model is capable to reproduce the behavior we can observe in some empirical ecosystems, e.g., the Glanville fritillary butterfly \emph{Melitaea cinxia} in the \r{A}land islands and the giant kelp \emph{Macrocystis pyrifera} in southern California \cite{,hanski1994metapopulation,orsini2008genetic,hanski2011eco,hanski2017ecological,opedal2020host,castorani2015connectivity,bell2015remote}. In both cases, the landscape is naturally fragmented in several available habitat patches, however empirical observations show that most of them remain empty because animals spontaneously select a subset of them where to live. 

\JFadd{In the fashion of neutral theory we here consider a single species model (or of indistinguishable individuals)} capable to reproduce the emergence of empty niches hereby strictly intended as vacant (local) habitats following Grinnellian formalism \cite{grinnell1917niche}, and the resulting fragmentation of the \JFadd{available habitat into smaller isolated ones} \cite{goodnight_evolution_2011, wilson_complex_1992}. \teoadd{To this aim, we focus exclusively on a dispersal process, deliberately neglecting other types of dynamics (e.g., birth/death processes). Doing so, we assume that any aspatial intra- or interspecific dynamics occur considerably faster than the spatial one. Furthermore, such local dynamics are assumed not to be affected by exogenous factors, thus overall conserving the total mass of the species involved (see Sec. \ref{sec:conclusions} for further details).} The model we propose can help \teoadd{to} better understand the potential mechanisms that lay the ground for phenomena such as peripatric speciation or endogenous habitat fragmentation. It is important to emphasize that in the latter case, the fragmentation occurs exclusively due to the population dynamics of the species rather than exogenous factors that influence the habitat landscape. %From this perspective, the concept of empty niche, hereby intended in the Grinnellian sense \cite{grinnell1917niche}, is an accepeted outcome of our formalism.} 
Our approach is reminiscent of the celebrated Schelling's model \cite{schelling_1969, schelling_dynamic_1971, rogers_2011_a}, according to which self-organised patterns of segregation spontaneously emerge due to the interactions between (at least) two antagonistic types of individuals. At variance with such paradigm, and in the spirit of neutral theory, we hereby assume all \teoadd{individuals} to be equivalent (neutral paradigm) \cite{leibold2004metacommunity}. Within this framework, we show that \teoadd{the fragmentation of the habitat} is now exclusively due to the spatial heterogeneity of the networked landscape, particularly the presence of multiple cores linked together through bridge (peripheral) nodes. \teoadd{The paper is organised as follows\JFadd{:} in Sec. \ref{sec:model}, we introduce the model and derive the \teoadd{mean-field} equations describing the dispersal of individuals across the patches. \teoadd{We investigate} the long-term distribution of the population sizes and the emergence of vacant patches \teoadd{in Sec. \ref{sec:fixed_points}}. In Sec. \ref{sec:topology}, \teoadd{we illustrate} our results using real and synthetic networks of habitat patches \teoadd{as a substrate of the dynamical process. In particular, we show how our model can result useful in understanding the habitat fragmentation of \emph{Glanville fritillary} butterfly, a species endemic of \r{A}land islands}. We then conclude \teoadd{and further discuss the perspective impact} of our work.} %\JFcomment{I moved the last paragraph of the introduction to the next section describing the model.} \\

\section{Self-segregation process}\label{sec:model}

\teoadd{We} consider a metapopulation network, whose nodes are supposed to represent the existing spatial patches or the local habitats connected through edges \teoadd{mimicking physical paths bridging distant patches}. In our model, the only random process is the dispersal of individuals, \JFadd{assumed to be indistinguishable}, driven by the affinity that individuals of different but adjacent nodes exhibit for each other. \JFadd{We thus assume an overall carrying capacity that sets the total population size at an (average) fixed value. As anticipated earlier, this assumption holds true when the timescale of the local dynamics on each of the spatial patches is disentangled from the global one of the dispersal across the different patches \cite{asllani2020dynamics}.} \textcolor{black}{For instance, if we assume the local birth/death processes to be very slow, the total population can be safely assumed to be constant during the whole diffusion process. As we will illustrate in the following through an empirical example, another plausible alternative is to consider that such local dynamics are much faster than the dispersal one. Indeed, the last case can allow splitting the dynamics of the model into two steps, the first (very short) phase where the local populations reach a stationary value in each node, followed by a second where individuals move among the habitats.} \teoadd{The} interactions \teoadd{among individuals of adjacent patches} are grounded on the mutual benefit that individuals have when sharing the same biogeographical area or territory. Examples span from more extreme cases such as the clumping of individuals of the same species to prevent the Allee effect \cite{allee_studies_1932, jorgensen_encyclopedia_2008} to less drastic ones such as the sharing of habitat of territorial animals during the mating season \cite{hixon_territory_1987}. We will denote by \emph{self-\teoadd{segregation}} the positive influence that the \teoadd{indistinguishable} individuals exert on each other. On the other hand, to model/introduce the habitat saturation conditioned by limited (abiotic) resources or negative interactions such as intra- or inter-species competition, etc., we assume that the nodes have a maximum carrying capacity of possible individuals to allocate. Hence, populated spatial patches are highly likely to attract new agents as long as they do not become too crowded. The resulting collective behavior is the emergence of different configurations of \JFadd{isolated (sub)communities } \JFadd{- each (sub)community consisting of a set of connected and occupied spatial patches -} surrounded by vacant patches of the habitat. \textcolor{black}{The terminology \textit{(sub)community}, as used here, is borrowed from network science \cite{newman_2017_networks}, where it emphasizes the presence of strongly connected (sub)networks clearly distinguished from each other. Let us nevertheless mention that an alternative definition in ecology would be a separated/isolated (sub)population \cite{leibold2004metacommunity}. However, for reasons of wording compactness, throughout this paper, we have preferred to use the term \textit{(sub)community} instead.}

We start by considering a simple {(i.e., without self-loops or multi-edges)} connected and undirected graph made of $\Omega$ nodes $\{v_i\}_\Omega$ whose structure is encoded in its adjacency matrix entries, $A_{ij}=1$ if there is a link between nodes $v_i$ and $v_j$, and zero otherwise. A fundamental assumption of our model is that the network is characterised by a heterogeneous degree distribution $\{p(k_i)\}_\Omega$ where the degree of node $v_i$ is defined as $k_i=\sum_{j=1}^\Omega A_{ij}$. \JFadd{The population size at time $t$ within node $v_i$ is written $n_i(t)$.}
If we denote the state of the system at time $t$ by $\mathbf{n}(t)\triangleq(n_1(t),n_2(t),\cdots,n_\Omega)$ and the probability of observing such state by $P(\mathbf{n},t)$, then the mathematical formalism that governs the individuals dynamics is described by the master equation \cite{gardiner_handbook_2004, kampen_stochastic_2007}:
\begin{equation}
\frac{\text{d} P(\mathbf{n},t)}{\text{d}t} = \sum_{\mathbf{n'}\neq \mathbf{n}}\Big[
T(\mathbf{n}\vert\mathbf{n'}) P(\mathbf{n'},t) - T(\mathbf{n'}\vert\mathbf{n}) P(\mathbf{n},t)
\Big],\,
\label{eq:master_eq}
\end{equation}
where $T(\mathbf{n'}\vert\mathbf{n})$ stands for the transition probability from state $\mathbf{n}$ to state $\mathbf{n'}$. Following the idea of the finite carrying capacity of the habitat patches (and similarly to \cite{asllani2018hopping,carletti_2020_nonlinear}), we impose to each node $v_i$ a maximum number of agents $1\leq n_i\leq N$ that can be hosted. Also to focus exclusively on the effect of the attractiveness that agents exert on each other and at the same time their intention to avoid overcrowding, we assume that individuals are not allowed for other actions. With a slight abuse of notation, the transition from node $v_i$ to  node $v_j$ is written as:
\begin{equation}
T(n_i-1,n_j+1\vert n_i, n_j) = \frac{A_{ij}}{k_i}\frac{n_i}{N} g\left(\frac{n_j}{N}\right),
\end{equation}
where the function $g(\cdot)$ represents the probability for the agents to settle in the chosen host node and quantifies the \teoadd{reciprocal benefit} among individuals while still considering the finite size of the nodes. A significant difference with other random processes, e.g., the biased random walks \cite{fronczak2009biased}, is that the probability of choosing the host node is independent of the densities of individuals in the other neighbour nodes \footnotetext[1]{In a biased random walk, the motion of individuals depends on the relative attribute of the hosting node compared to all the neighbour ones \cite{fronczak2009biased, gmezgardees_2008_entropy}.} \cite{Note1}. This subtle but crucial feature makes sense for the spatial interpretation of the habitat support \cite{fanelli_diffusion_2010, alonso_merits_2006}: the individual cannot \textit{a priori} choose the most suitable node before first ``testing the ground''. Said differently, the agents perceive the pressure of limited resources or other individuals of the same kind once they reside on a given patch, but not earlier.  
Throughout this paper, we will make a straightforward assumption regarding the function $g(\cdot)$: the hosting node will recruit individuals proportionally to the density of agents already present therein, and at the same time, the probability for a recruited individual to settle in the selected node, is proportional to the available free space (capacity constraint). In formula we have
\begin{equation}
g\left(\frac{n_j}{N}\right)=\frac{n_j}{N}\frac{N-n_j}{N}, \nonumber
\end{equation}
which constitutes the prominent logistic function and has plentiful applications in many areas of science and in particular in ecology \cite{murray_mathematical_2002}. 
\teoadd{Since looking for an exact solution of Eq. \eqref{eq:master_eq} results challenging in general, we apply} a standard procedure \teoadd{to obtain the deterministic description of the self-segregation problem.} We \teoadd{start by} multiplying both sides of Eq.~\eqref{eq:master_eq} by $n_i$, and summing over $n_i$ for all $i$, we obtain the evolution equation for the average density \teoadd{\[\langle n_i(t)\rangle=\sum_{\mathbf{n}}n_i P(\mathbf{n},t)=\sum_{n_i}n_i\sum_{\mathbf{n}\setminus n_i}P(\mathbf{n},t)=\sum_{n_i}n_iP(n_i,t)\] recalling that $P(\mathbf{n},t)$ is the joint probability distribution $P(n_1,\dots,n_{\Omega},t)$. Regarding the r.h.s. of Eq. \eqref{eq:master_eq} instead, we need to point out that the only available state reachable from $\mathbf{n}(t)=(n_1,n_2,\cdots,n_i,\dots, n_j,\dots,n_\Omega)$ in the time interval $\Delta t$ is $\mathbf{n}(t)=(n_1,n_2,\cdots,n_i\mp 1,\dots, n_j\pm 1,\dots,n_\Omega)$ where $v_i$ and $v_j$ are two adjacent nodes. Thus by shifting the two indices of the two sums of Eq. \eqref{eq:master_eq} respectively by $+1$ and $-1$ we obtain \[\frac{\text{d} \langle n_i\rangle}{\text{d}t}=\sum_{j=1}^\Omega \langle T(n_i+1,n_j-1|n_i,n_j)\rangle - \langle T(n_i-1,n_j+1|n_i,n_j)\rangle \]} \teoadd{where for sake of simplicity we have ommitted the inactive terms $n_l$ and where again $\langle f(\mathbf{n},t)\rangle=\sum_{\mathbf{n}}f(\mathbf{n},t) P(\mathbf{n},t) $}. \teoadd{To obtain the deterministic equation, we proceed by dividing both sides by the carrying capacity $N$ and rescale the time $t \mapsto t/N$.} \teoadd{The} thermodynamic limit ${N\rightarrow+\infty}$ yields the mean-field (MF) equations:  
\begin{equation}
\frac{\text{d} \rho_i}{\text{d}t}=\sum_{j=1}^\Omega\mathcal{L}_{ij}\rho_i\rho_j\left[1-\rho_i - \dfrac{k_j}{k_i}(1-\rho_j)\right],\, \forall i,
\label{eq:MF0}
\end{equation}
where $\rho_i\equiv\lim_{N\rightarrow+\infty}\langle n_i\rangle/N$ is the \JFadd{population density within node $v_i$} and  $\mathcal{L}_{ij}=A_{ij}/k_j-\delta_{ij}$ corresponds to the random walk (RW) Laplacian \cite{newman_2017_networks,barrat2008dynamical}. Notice also that in the limit of large $N$ we drop any correlation among different nodes i.e., $\langle n_i n_j \rangle \sim \langle n_i \rangle \langle n_j \rangle$ based on the van Kampen {ansatz} \footnotetext[2]{\teoadd{Based on van Kampen ansatz, we can write the (rescaled) discrete variable as $n_i/N=\rho_i + \xi_i/\sqrt{N}$, valid for large $N$, where $\rho_i$ is the deterministic variable describing the density of node $v_i$ and $\xi_i$ the stochastic variable. Thus $\langle n_in_j\rangle/N^2=\langle \rho_i\rho_j + \xi_i\xi_j/N+\left(\xi_j\rho_i+\xi_i\rho_j\right)/\sqrt{N}\rangle\sim\langle n_i\rangle \langle n_j \rangle/N^2$ in the limit for large $N$ while the densities are kept constant.}} \cite{kampen_stochastic_2007,Note2}. 
\JFadd{From Eq. (\ref{eq:MF0}), it can be easily shown that the total population size is constant over time, i.e., $\sum_{i}\frac{\text{d} \rho_i}{\text{d}t} = 0$, a consequence of the fact that only dispersal processes are considered. We will define the average density $\beta=\sum_{i} \rho_i/ \Omega$ taking values in the interval $[0,1]$. }	

\begin{figure*}[t!]
\centering
\includegraphics[width=\textwidth]{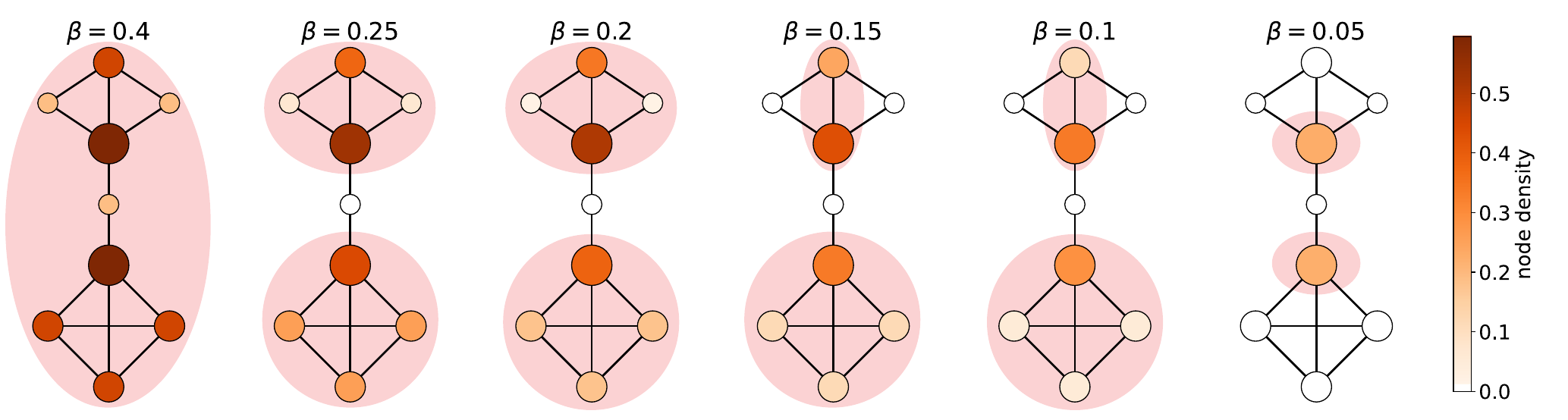}
\caption{Self-\teoadd{segregation} and the ensuing \JFadd{(sub)communities} in a heterogeneous network. In all the panels, nodes size and colour represent, respectively, their degree and the asymptotic mass density (the larger the size the higher the degree and the darker the colour the larger the density); moreover each system has been initialised with a uniform density of individuals per node, i.e. $\rho_i(0)=\beta$. When the total mass is sufficiently large (leftmost panel), all the nodes are occupied and a single large community prevails (emphasised with the shaded oval). However, once the average density $\beta$ starts decreasing (from the left panel to the right one), nodes are depleted accordingly to their degree, the smallest the first, leaving only nodes with higher degrees occupied. Among the nodes of degree $2$, the first one to become empty is the one with the largest bridgeness \cite{jensen2016detecting} (see Appendix \ref{App_B}). In such a scenario, (connected) subgraphs with filled nodes give birth to separated (sub)communities (emphasised with the shaded ovals) separated by of empty nodes. \JFadd{For instance, when $\beta=0.15$, the mass segregates into two subcommunities $\mathcal{C}_1$ ad $\mathcal{C}_2$ of respective sizes $4$ and $2$. In the subcommunity $\mathcal{C}_2$, the two nodes have distinct stationary densities since they have different degrees.} Notice also that while the average density $\beta$ varies, the number of vacant nodes does not necessarily change accordingly, a phenomenon we refer to as ``quantisation'' of the occupancy (see Fig.~\ref{fig:nested} for further details).}
\label{fig:metacommunities}
\end{figure*}

\section{Fragmentation, multistability and resilience of (sub)communities}\label{sec:fixed_points}

Starting from the \JFadd{mean-field Eq.}~\eqref{eq:MF0} we can determine the equilibrium states $\{\rho_i^*\}_\Omega$ and based on that we can afterwards perform a (linear) stability analysis. \JFadd{Let us first note that any node might potentially become vacant as $\rho_i^* = 0$ is a fixed point of system (\ref{eq:MF0}) $(i=1,\cdots,\Omega)$. Obviously, there is a limited number of nodes that can become vacant as the total mass must remain constant while satisfying the limiting carrying capacities of the nodes. As empty nodes emerge, the mass can segregate into $M$ isolated subcommunities, the $m$-th ($m=1,\cdots,M$) one being a subnetwork of $\Omega_m$ connected and filled nodes. Such subcommunities are separated by empty patches that prevent individuals from migrating to another subcommunity.}
A straightforward computation (the interested reader can refer to the Appendix \ref{App_A}) allows to obtain the \JFadd{population density} in the \JFadd{$m$-th one} at equilibrium to be given by 
\begin{equation}
\rho_i^*=0\qquad \mathrm{or} \qquad \rho_i^*=1-\frac{C_m}{k_i},\, \forall i,
\label{eq:stat_rel}    
\end{equation}
where $C_m=\left(1-\beta_m\right)/\langle 1/ k\rangle_{\Omega_m}$ is a conservation constant that depends on the initial configuration $\{ \rho_i(0)\}_\Omega$ and {$\beta_m=\sum_{j\in M_m}\rho_{j}^*/\Omega_m$} is the average {stationary} \JFadd{population density} of the $m$-th \JFadd{subcommunity} {$M_m$ with size $\Omega_m$}. \JFadd{Notice that, in general, the interpretation of the system behavior depends on the knowledge of the conservation constant $C_m$ which cannot be a priori inferred.} Eq.~\eqref{eq:stat_rel} results thus an implicit equation for the stationary nodes' densities. \teoadd{Let us point out also that in equation (\ref{eq:stat_rel}), $k_i$ is the degree of node $v_i$ considering the whole network, not only the subcommunity to which it belongs.}
In particular, in the case of a single \JFadd{community}, $\beta_1 = \beta$ is the {global density, i.e., the} total number of agents divided by the number of  {non empty} nodes. Observe that the average $\langle 1/ k\rangle_{\Omega_m}$, has been performed over the $\Omega_m$ nodes forming the \JFadd{$m$-th subcommunity}.
From relation \eqref{eq:stat_rel}, it is straightforward that nodes resulting occupied at equilibrium should have a degree $k_i>C_m$ or will be vacant otherwise. Because $C_m$ decreases with the total mass, the last observation suggests that acting on the mass we can induce the agents to segregate at the nodes with higher degrees, possibly leaving the ones with a lower degree vacant. 
To elucidate the robustness of  the \JFadd{(sub)communities emergence} we will consider the case when a single empty node, adjacent to a given \JFadd{(sub)community}, is slightly perturbed \footnotetext[3]{Obviously, an isolated, empty node (surrounded by other empty ones) is always unstable since the mass added to it will never leave the node.}\cite{Note3} and postpone a complete stability analysis to Appendix \ref{App_A}. We start by assuming the system settles on some equilibrium and we focus on the $m$-th \JFadd{subcommunity}, whose nodes are characterised by $\rho_j^*=1-C_m/k_j>0$. Suppose it exists an empty node $v_i$ connected exclusively to the \JFadd{nodes of the $m$-th subcommunity} to which we allocate a small amount of mass, such that this new configuration still satisfies the same constraints {($m$ and $\beta_m$ remain unchanged assuming $\Omega_m$ is large enough to neglect the small amount of mass we have to remove from each node but the $i$--th one to satisfy the constraint)}, that is $\rho_i^*\mapsto \delta$ and $\rho_j^*=1-C_m/k_j$ for $j\neq i$. A direct computation allows to obtain the linear{ised} dynamics that governs node $v_i$ at early times
\begin{equation}
\frac{\text{d} \delta}{\text{d}t}\approx \delta \left(1-\frac{C_m}{k_i}\right)\sum_{\substack{j=1\\\left(j\neq i\right)}}^{\Omega_m}\mathcal{L}_{ij}\rho_j^* \,.
\label{eq:lin3}
\end{equation}
From here we can conclude that the $v_i$-th node will increase its mass (acquiring it from other $\Omega_m$ nodes) if $C_m<k_i$ and will transfer it to the other nodes of the (sub)community $m$ otherwise. Similarly, if we add a small mass $\delta$ to an occupied node, the system is always stable. In fact, if we slightly perturb only the $i$-th node of the \JFadd{$m$-th subcommunity}, $\rho_i^*\mapsto \rho_i^*+\delta$, with $\delta>0$, a straightforward computation allows writing $\frac{\text{d} \delta}{\text{d}t}\approx -\delta \rho_i^*\sum_{j=1}^\Omega\mathcal{L}_{ij}\rho_j^*\, ,$ thus proving the stability of the \JFadd{subcommunity}. This is reasonable since the added mass to some pre-existing \JFadd{(sub)community} cannot escape from it being surrounded by empty nodes \teoadd{acting as movement barriers}.

The local analysis emphasises the role of the degree distribution heterogeneity in the existence of stable states with empty nodes. For instance, in a regular graph, the only possible (non-trivial) stable state is the one being uniformly occupied since $C_m<k$ for all the nodes \footnotetext[4]{This can be easily obtained from the formula $C_m=\left(1-\beta_m\right)k$ since the average $\langle \cdot\rangle_{\Omega}$ over the degrees drops being the graph regular. From here, it is immediate that $C_m/k=\left(1-\beta_m\right) < 1$, which justifies the stability of the fully occupied nodes state.}\cite{Note4}. %\JFdel{Nevertheless, in the following, we will see that, for non-regular networks, a node can be either empty or occupied depending on the combination of a random choice of the initial conditions and the structural features of the network structure. The latter yields to {stable configurations with different numbers \JFadd{and distributions} of vacant nodes \JFadd{even for a fixed value of the metapopulation size} }, a phenomenon we hereby refer to as multistability.} 
To illustrate the outcomes of the self-\teoadd{segregation} dynamics, we refer to Fig.~\ref{fig:metacommunities}, where the system has been initialised with a uniform distribution of nodes' densities for all the panels. From the previous analysis, if the average density is sufficiently large, i.e., $\beta > 1-{1}/{\Omega}$, all nodes are occupied at equilibrium, resulting in a single \teoadd{community}. However, as $\beta$ is gradually reduced, some empty nodes (the least connected ones {and those with the largest bridgeness centrality \cite{jensen2016detecting}}) start to emerge. Decreasing the total mass would result in increasing the constant $C_m$, which in turn forces all the nodes for which $k_j>C_m$ is no longer satisfied, to deplete. The mass accumulates on fewer nodes in such a way to satisfy the new equilibrium condition~\eqref{eq:stat_rel}. Also from Fig.~\ref{fig:metacommunities}, one can notice that {for (sufficiently) small variations of} the mass, the set of empty/occupied {nodes} remains the same, showing that the system is resilient in switching to a state with more (less) vacant nodes while the mass decreases (increases). 
Such ``quantisation'' phenomenon is exclusively due to the discrete distribution of the network support and can be understood starting from the conservation constant $C_m=\left(1-\beta_m\right)/\langle 1/ k\rangle_{\Omega_m}$ which changes continuously with $\beta_m${, if the number of \teoadd{(sub)communities and empty patches} does not vary,} that in turn depends on the total average density $\beta$. Since the degrees are discrete and the network is finite, density intervals, where $C_m$ does not overtake the next lowest degree of occupied nodes, will certainly exist. Furthermore, the length of such intervals must decrease when the differences between the successive degrees become smaller, suggesting that a broader degree distribution will show a less pronounced quantisation effect.
\JFadd{Nevertheless, in the following, we will see that, for non-regular networks, a node can be either empty or occupied depending both on the initial conditions and the network structural features. The latter yield to stable configurations with different numbers and distributions of vacant nodes even for a fixed value of the metapopulation size, a phenomenon we hereby refer to as \emph{multistability}. The latter yield stable configurations with different numbers and distributions of vacant nodes for a fixed value of the metapopulation size, a phenomenon we hereby refer to as multistability. More precisely, we will show the existence of regions of the phase space, determined by $\beta$, i.e., the hyperplane $\beta_\Omega = \sum_i \rho_i(0)$, containing several attractors, each one with a relatively small stability basin thus even slightly changed initial conditions can determine very different system outcomes, in particular with a substantial change in the number of empty nodes.}\\

\section{Role of topology in the formation of (sub)communities}\label{sec:topology}

\teoadd{So far, we have shown that the heterogeneous degree distribution characterizing the network supports is crucial in the fragmentation of the habitat in isolated (sub)communities separated by empty patches. In this section, we explore in detail the structural features responsible for the breaking of the contiguous habitat and the emergence of vacant ones in both synthetic and empirical ecological networks. In particular, we show that low degree bridge nodes are responsible for the breaking of the original community in smaller ones by being the first to deplete. Let us remember the definition of bridge nodes, the latter being nodes with high betweenness centrality and connecting different and distant regions in the network. They thus act as global bridges allowing separate parts of the network to communicate, in opposition with local bridges that connect nearby groups of nodes. Removal of global bridges will thus easily disconnect the network. In the next two sections we will consider the behavior of the above defined model on a synthetic spatial network, the random geometric network, and on the empirical network arising from the ecosystem of the Glanville fritillary butterfly in the \r{A}land islands. We further investigate the role of network topology on the metapopulation breaking for other synthetic and real networks in Appendix \ref{App_B}.}

\JFadd{\subsection{Random geometric networks}}

\begin{figure*}
		\centering
		\begin{tabular}{cccccc}
			\includegraphics[width=.3\linewidth]{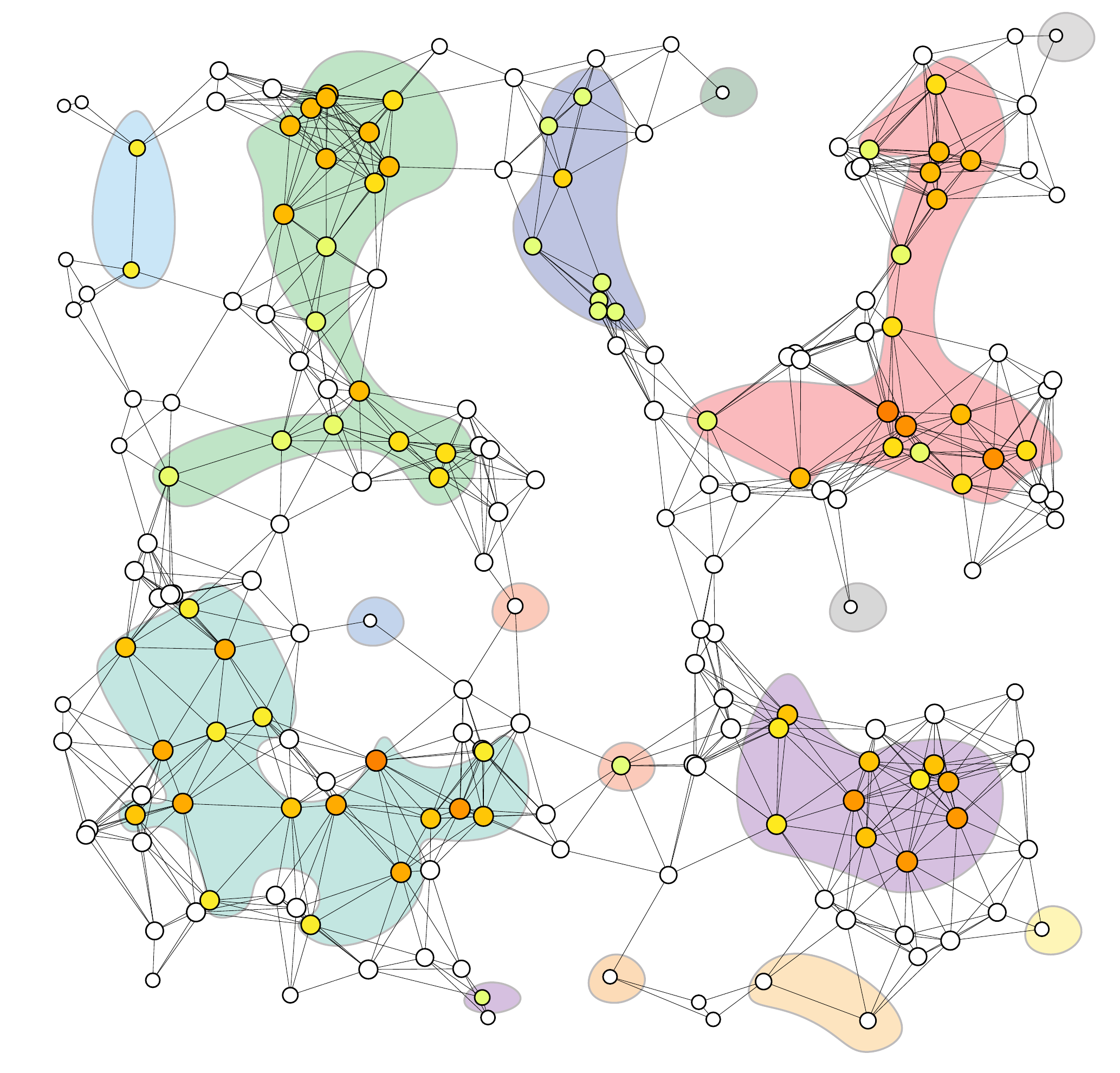}&
			\includegraphics[width=.3\linewidth]{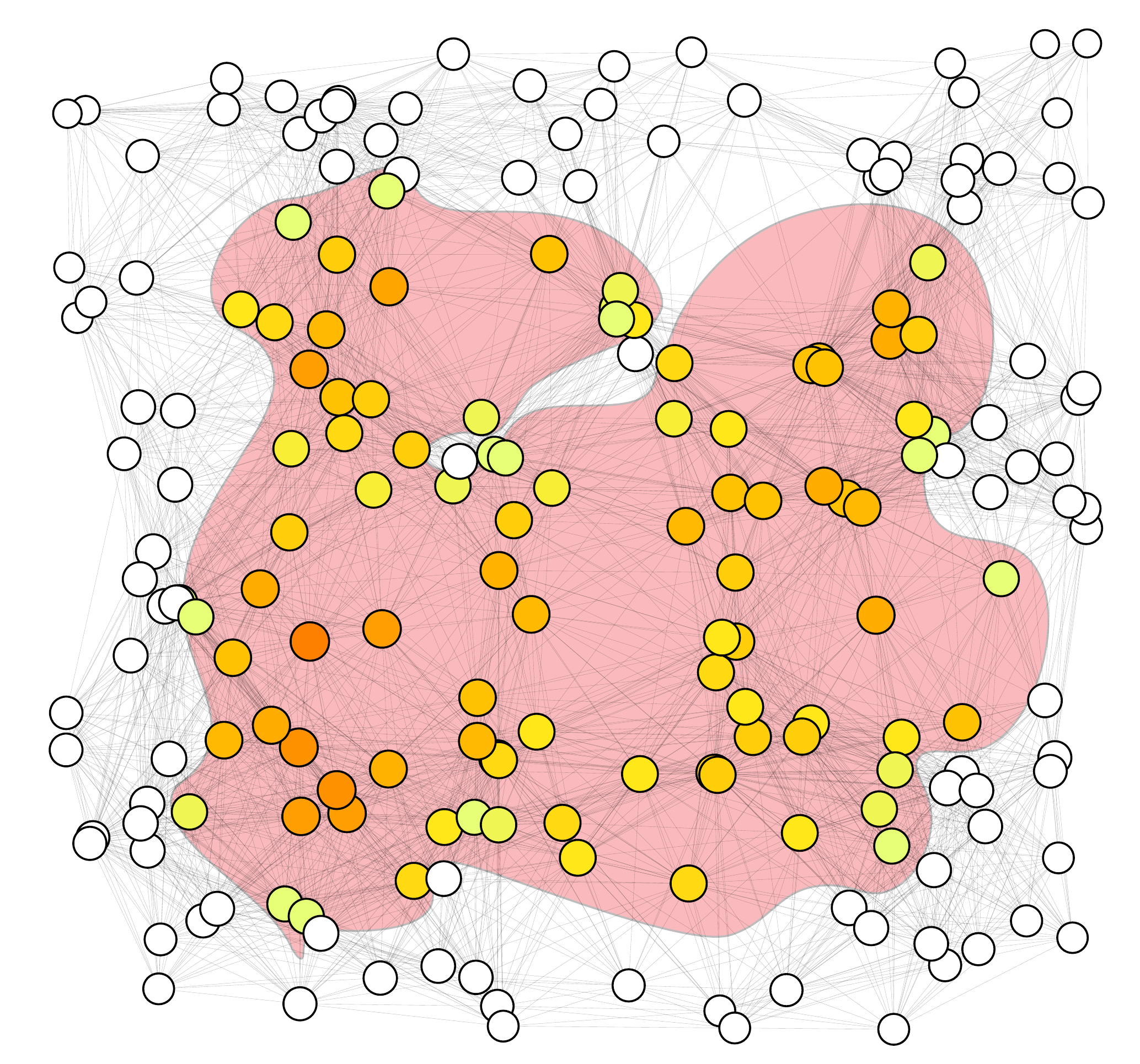}&
			\includegraphics[width=.3\linewidth]{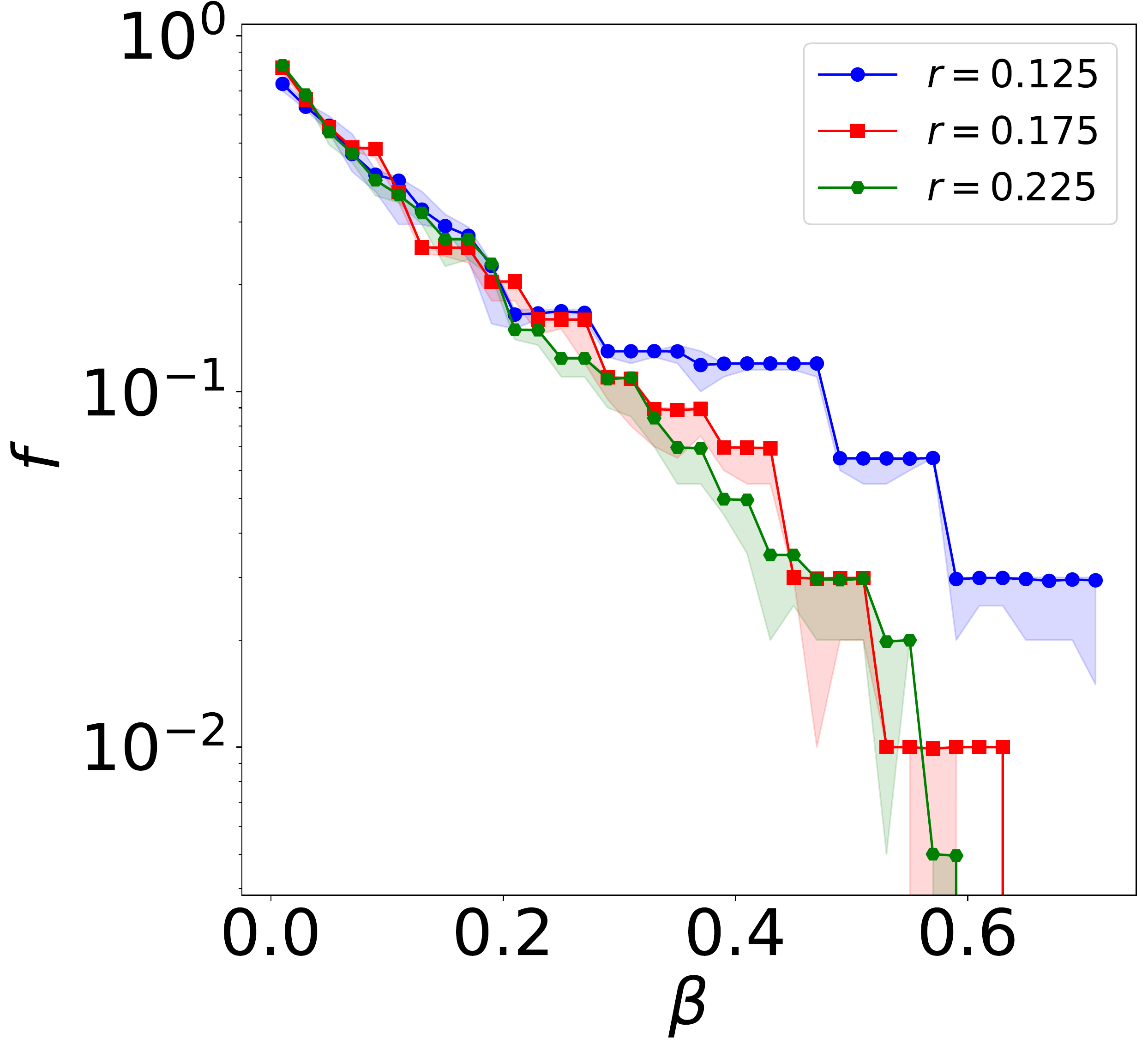}\\
			(a) & (b) & (c)
		\end{tabular}
		
		\caption{ \JFadd{(a) Emergence of isolated (sub)communities separated by empty nodes in a random geometric graph made of $\Omega=200$ nodes randomly drawn in the unit square. Every pair of nodes is connected if their Euclidean distance is less than $r=0.175$. White nodes correspond to empty nodes while colored nodes correspond to filled ones. Each shaded area corresponds to a subset of connected filled nodes. (b) The same but for $r=0.275$. (c) Fraction of empty nodes in a random geometric graph made of $\Omega=200$ nodes, as a function of $\beta$. Each curve corresponds to a distinct value of $r$.}}
		\label{fig:RGG}
\end{figure*}

\JFadd{We start by considering a family of random geometric graphs as a test bed for numerically verifying the theory developed throughout this paper. Such spatial networks have resulted useful for modelling the dispersal of individuals in ecological settings \cite{grilli2015metapopulation,gross2020modern,ryser2019biggest}. Starting from a set of uniformly distributed nodes in the Cartesian space, we connect every two nodes whose Euclidian distance is lower than some threshold, the rationale being that individuals are more likely to migrate to spatial patches that are not too far away. Panels $a)$ and $b)$ of Fig. \ref{fig:RGG} correspond to random geometric (connected) graphs obtained by drawing $\Omega=200$ points in the unit square and connecting them if their euclidean distance is lower than $r=0.175$ or $r=0.275$ for panel $a)$ and $b)$, respectively. This procedure results in two distinct graphs on top of which we ran the dynamical system given by (\ref{eq:MF0}), considering an average node population density $\beta=0.05$. In both cases, we emphasize the distinct emerged (sub)communities by means of shaded areas. While a single giant community is observed when $r=0.275$, multiple isolated (sub)communities are found when $r=0.175$. In panel $c)$, we show the fraction of empty nodes as a function of $\beta$, for random geometric graphs with distinct values of the parameter $r$. We observe a quantisation effect in the fraction of empty nodes. As $r$ increases, this quantisation effect becomes more diluted since the degree distribution becomes sharper. The results were averaged over different configuration of the initial distributions of the densities $\rho_i(0)$, and the min-max deviation is reported by the shaded area. Hence, the system is multistable, namely for the same value of $\beta$ and a fixed network, the asymptotic distribution of mass can vary because it depends on the initial conditions.\newline}

\subsection{The Glanville fritillary metapopulation}

\begin{figure*}
	\centering
	\begin{tabular}{cccccc}
		\includegraphics[width=.35\linewidth]{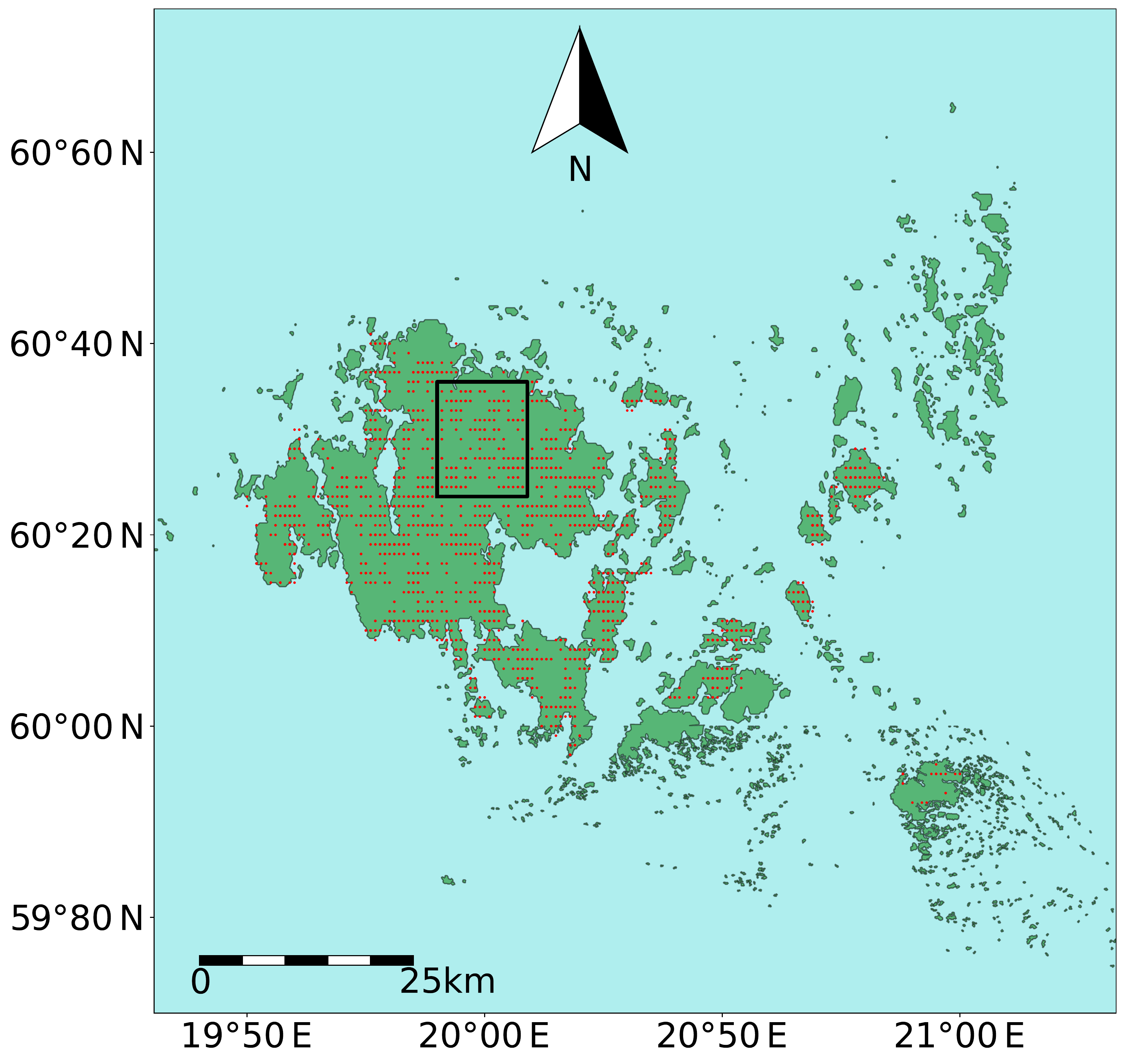} &
		\includegraphics[width=.35\linewidth]{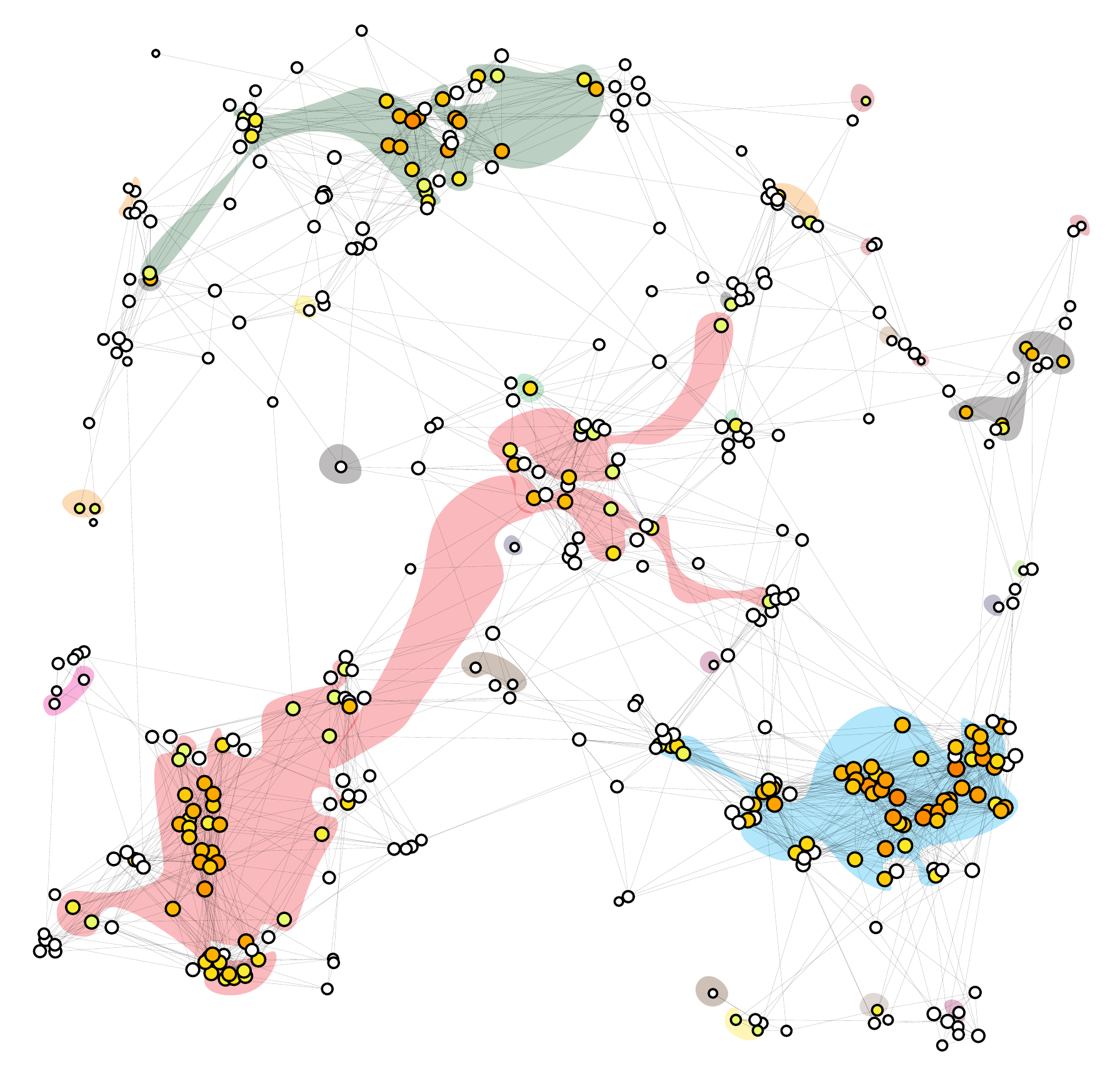}\\
		(a)&(b)\\
		\includegraphics[width=.35\linewidth]{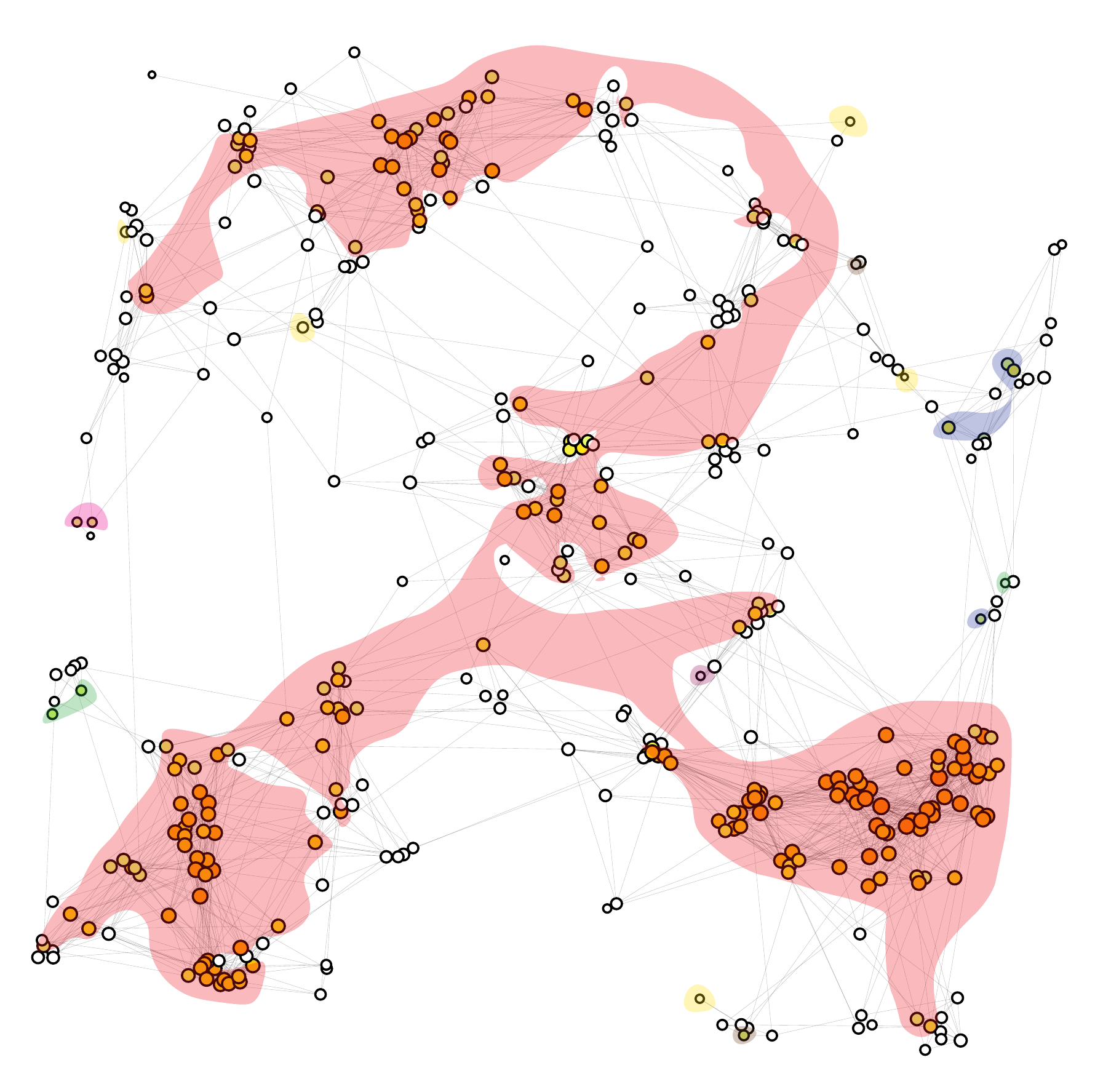}&
		\includegraphics[width=.35\linewidth]{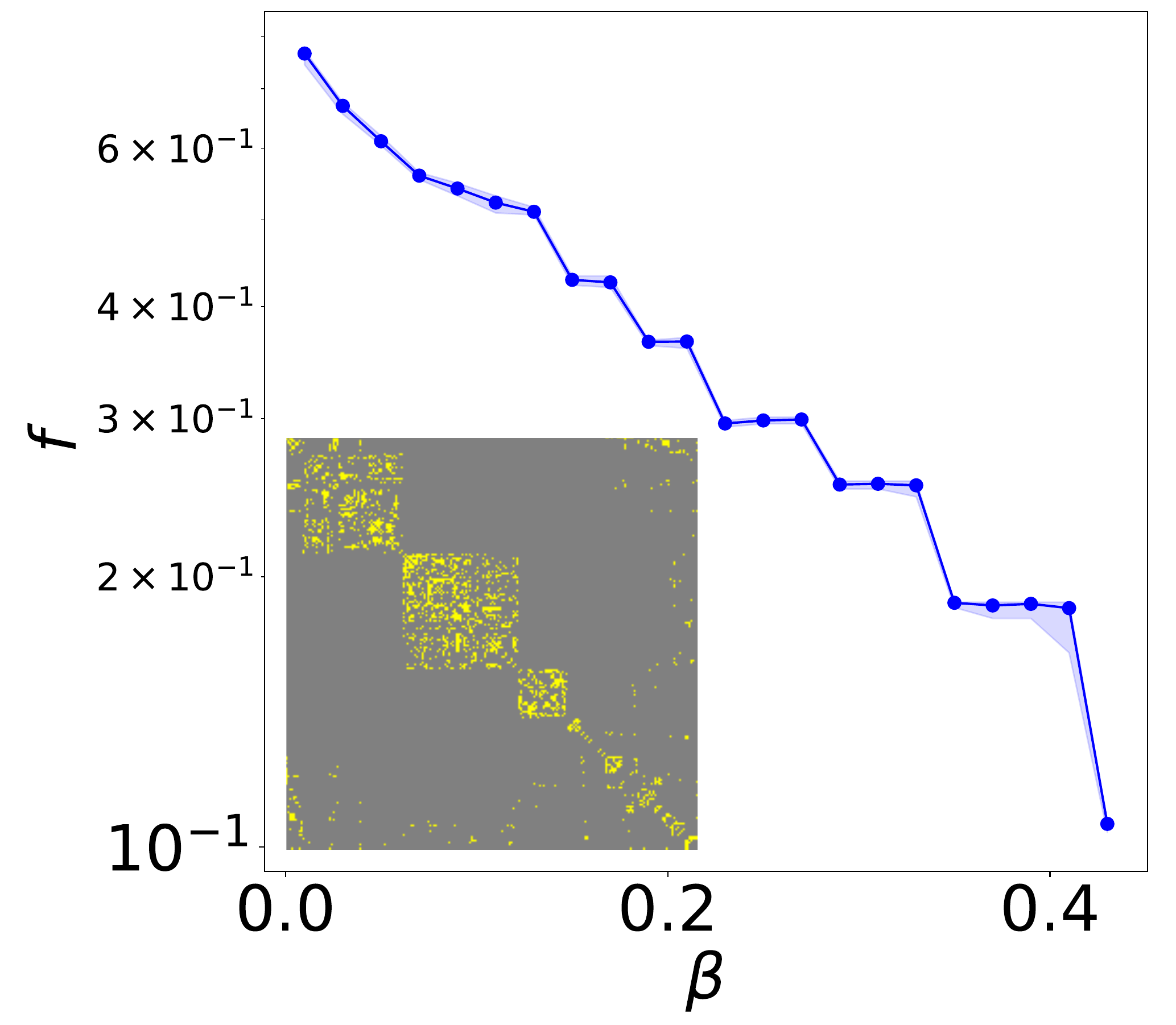}\\
		(c) & (d)
	\end{tabular}
	\caption{ \JFadd{ $(a)$ Map of the \r{A}land islands (southwestern Finland). The boundary of \r{A}land islands was retrieved from \cite{mapAland}. The suitable habitat patches \cite{butterfly_patches} of the Glanville fritillary butterfly are indicated by the red points. We here restrict our attention to the subset of $395$ habitat patches found within the black rectangle. $(b)$ Emergence of isolated (sub)communities (shaded areas) separated by empty patches (white nodes) for $\beta=0.08$ in the random geometric graph corresponding to the subset of habitat patches (see the main text for the construction of this graph). $(c)$ The same but for $\beta = 0.14$. In this case, there is a single large community in addition to a few smaller ones. $(d)$ Fraction of empty nodes as a function of $\beta$ with the min-max deviation. The inset highlights the core-periphery structure of the graph showing the adjacency matrix.}}
	\label{fig:butterfly}
\end{figure*}

\JFadd{In this section, we will switch our attention to an empirical spatial ecological network representing the natural habitat of the \textit{Melitaea cinxia} species. Commonly known as Glanville fritillary butterfly, it is an endemic species of the \r{A}land islands in southwestern Finland \cite{hanski2011eco}. Its habitat consists of a fragmented landscape of meadows, pastures and rocky coastal areas separated one from the other by uninhabitable areas \cite{opedal2020host}. Butterflies disperse from patch to patch with an average migration distance of a few hundred meters to a few kilometers \cite{opedal2020host,hanski1994metapopulation,hanski2017ecological}. Interestingly, an important fraction ($75-80 \%$) of these suitable habitat patches are left vacant, i.e., are unoccupied by the butterflies \cite{opedal2020host}. Although birth and death processes as well as extinction and colonisations events occur at the level of the patches, the size of the metapopulation has remained relatively stable over longer timescales, supporting the idea of a fixed global carrying capacity \cite{orsini2008genetic,hanski2011eco}. So far, around $4500$ habitat patches suitable for hosting this butterfly species have been catalogued \cite{opedal2020host}. For the sake of clarity, we here restrict our attention to a subset of them found within the black rectangle shown in Fig. \ref{fig:butterfly} $(a)$. We draw connections between nodes assuming an exponential dispersal kernel; for every pair of distinct nodes, we connect them if their geodesic distance (in kilometers) is less than a random number drawn from the exponential distribution $f(x) = \lambda e^{-\lambda x}$, with $\lambda = 0.65$. This corresponds to an average dispersal distance of approximately $\dfrac{1}{\lambda}\approx 1.5$km, in adequation with biological data \cite{opedal2020host,hanski1994metapopulation,hanski2017ecological}. As expected, when $\beta$ is low enough, the mass segregates into numerous subnetworks separated by vacant patches, see panel $(b)$ obtained for $\beta = 0.08$. In this case, there are $222$ empty nodes and most of the mass accumulates into three (sub)communities of respective sizes $57, 56$ and $24$ nodes that correspond to the main cores of the network, as shown in the inset of panel $(d)$. Panel $(c)$ shows similar results for $\beta=0.14$, where this time a slightly greater total mass results in fewer  vacant nodes ($170$) and there is a single large community and few smaller ones. Panel $(d)$ shows the fraction of empty nodes as a function of $\beta$ with the min-max deviation.}

\JFadd{Let us mention that other species were also shown to live in a fragmented habitat and to disperse between the patches. This is for instance the case of African wild dogs in South Africa \cite{nicholson2020} or the giant kelp \emph{Macrocystis pyrifera} in southern California \cite{bell2015remote, castorani2015connectivity}.}

\section{Discussion and Conclusions}\label{sec:conclusions}

In this paper, we have introduced a self-\teoadd{segregation} process and have shown that it can induce a \JFadd{(meta)population network} to split in (sub)communities \teoadd{of contiguously occupied patches}, separated by empty \teoadd{ones}. \teoadd{We base our analysis on analytical study and numerical validation using both random geometric networks and empirical ecological ones.} The stationary process reveals the role of the network degree heterogeneity in the emergence of vacant nodes as novel states whose fraction is quantised during the node filling process and exhibits multistable configurations. Such features are absent in other random processes such as random walks~\cite{asllani2018hopping, carletti_2020_nonlinear,gmezgardees_2008_entropy,fronczak2009biased,lambiotte2014random,masuda_2017_random} or Schelling's segregation~\cite{schelling_1969,schelling_dynamic_1971,rogers_2011_a, gandica_can_2016,gargiulo_2017_emergent}. 
From the ecological perspective, the outcomes of our model stress that rich patterns endowed with vacant patches of the habitat can exist as a natural outcome of trade-off between first principles such as \teoadd{positive intra- or inter-species interactions e.g., cooperation and negative ones e.g.,} competition. Such theoretical evidence supports even more the vacant niche paradigm as a genuine consequence of population dynamics~\cite{kauffman_1993}. Furthermore, the quantisation effect illustrates the robustness of the occupancy configurations to changes in the {density} of individuals. The emergence of multistable states demonstrates that segregation can be affected by the randomness of the process, in line with the paradigm of the fitness landscape~\cite{kauffman_1993}. On the other side, our model proposes a robust mechanism for the fragmentation of the habitat in disconnected spatial patterns. We emphasize that such phenomenon is driven by endogenous factors, i.e., facilitations between individuals and most importantly by the structure of the \JFadd{metapopulation} network. Topological features of the spatial support have been recently shown to enable the patterning dynamics in minimal models of single species systems \cite{asllani_topological_2018}. Habitat fragementation has recently raised as a strong candidate for the speciations of animals or plants \cite{weir_ice_2004, dias_natural_2013, thompson_speciation_2018}. In particular, according to the peripatric speciation theory, the segregation of individuals in isolated macroscopic patches can promote the development of the new species as a consequence of isolated evolution \cite{coyne_speciation_2004}. 

\teoadd{The model we have proposed here focuses exclusively on a dispersal process which puts the emphasis on the spatial interactions between individuals in adjacent patches of the habitat rather than the local intrinsic dynamics of the species considered here, thus deliberatively neglecting the birth/death process or other intra- or inter-species interactions such as competition or mutualism. Such considerations hold true when the local and global timescales of the dynamics disentangle \cite{asllani2020dynamics} to allow a fast relaxation at the system's equilibrium at the patch level, consequently allowing overall conservation of mass by the slower segregation process. As an example and future potential extension of our model, one can consider the family of Lotka-Volterra models, which in the classic prey-predator case can produce periodic oscillations \cite{murray_mathematical_2002} or fixed point when a density-dependent is introduced \cite{kot2001}. However, as already emphasized, when such local dynamics tend to relax considerably faster to the equilibrium compared to the tendency to migrate, the total mass is kept constant on average, thus imposing an intrinsic fixed global carrying capacity.}

The relevance of vacant spatial patches can extend beyond the ecological setting and help better understand how segregation in human dynamics that has energetically flourished in many post-industrialised societies can lead to the spontaneous creation of urban prairies~\cite{prener_st._2020}. Similarly to what we have obtained in this work, a drop in the population density, usually triggered by deurbanization or high crime levels, has been responsible for the emergence of abandoned urban land that has reverted to green space consequently. Instead of taking advantage of the abundantly available cheap land, people have decided to segregate into (sub)communities where the population can enhance their fitness to better use of resources (e.g., infrastructure, public transport, education, etc.) \cite{prener_st._2020}. In this regard, we are confident that the segregation paradigm we present here can naturally extend from microscopic (bacteria, viruses, etc.) to macroscopic scale (e.g., human communities).

\section{Acknowledgments}

JFDK is supported by a FNRS Aspirant Fellowship under the Grant FC38477. Part of the results were obtained using the computational resources provided by the “Consortium des Equipements de Calcul Intensif” (CECI), funded by the Fonds de la Recherche Scientifique de Belgique (F.R.S.-FNRS) under Grant No. 2.5020.11 and by the Walloon Region.
\clearpage
\appendix

\section{The stability analysis of the fixed points}
\label{App_A}

\begin{figure}
\centering
\includegraphics[scale=1]{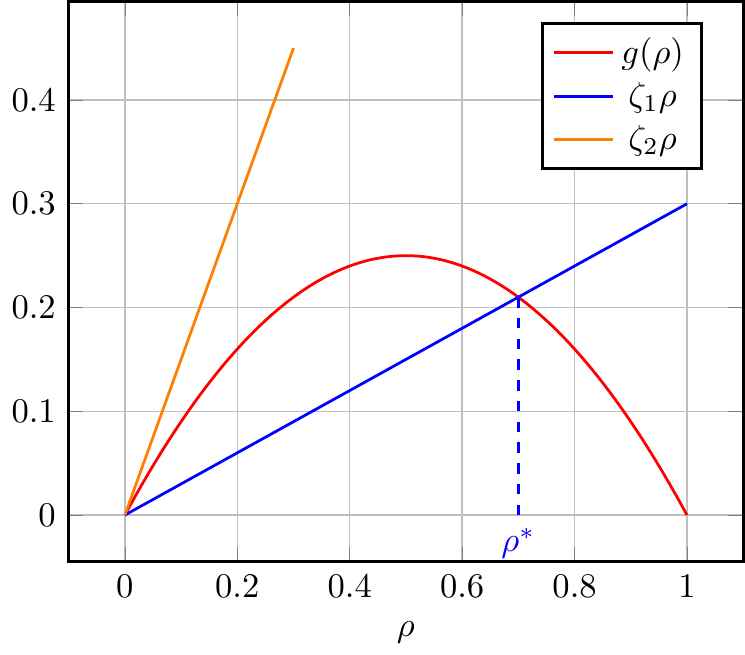}
\caption{  Geometrical interpretation of the stationary densities solution of Eq.~\eqref{eq:Stationary_density}. We show the graph of a generic function $g$ satisfying the required assumptions, together with two straight lines associated to two values of the node degree. {Either, there will be only one intersection point, corresponding to $\rho^* = 0$ or, for sufficiently large degrees, there can be two intersection points, namely $\rho^* = 0$ and $\rho^*>0$. In the latter case, only the non-zero solution is stable, as shown in Theorem $1$.}}
\label{fig:Geometrical_interpretation}
\end{figure}

We start by first emphasising that for the simple connected network made of $\Omega$ nodes the system of ODEs~\eqref{eq:MF} 
\begin{equation}
\frac{\text{d} \rho_i}{\text{d}t}=\sum_{l=1}^\Omega\mathcal{L}_{il}\rho_i\rho_l\left[1-\rho_i - \dfrac{k_l}{k_i}(1-\rho_l)\right],\, \forall i,
\label{eq:MF}
\end{equation}
of the main text generalises to:
\begin{align}
	\frac{\text{d} \rho_i}{\text{d}t}&=\sum_{l=1}^{\Omega}\mathcal{L}_{il}\Big[\rho_l g(\rho_i) - \frac{k_l}{k_i}\rho_ig(\rho_l)\Big]\nonumber\\
	&=\sum_{l=1}^{\Omega} A_{il}\Big[\f{\rho_l g(\rho_i)}{k_l} - \frac{\rho_ig(\rho_l)}{k_i}\Big],\, \forall i,
	\label{eq:MF1}
\end{align}
where we have kept a general form for the function $g(n_i/N)$. We assume $g$ to be a $C^1$ concave function on the open interval $(0,1)$ with $g(0)=0=g(1)$ and $g(x)>0$ for all $x\in(0,1)$. Consequently, the stationary density of node $v_i$ ($i=1,\cdots,\Omega$) is either $0$ or given by the unique non-zero solution (provided it exists) of the equation:
    \begin{equation}
	    \f{\rho_i^*}{C_{m} k_i}= g(\rho_i^*),
	    \label{eq:Stationary_density}
	\end{equation}
with the constant $C_{m}$ such that $\sum_{l\in M_{m}}\rho_l^*=\Omega_m\beta_{m}$ where the sum runs over the nodes belonging to the same community $M_m$ as node $v_i$ and $\Omega_m\beta_{m}$ is the total mass in this (sub)community.
Fig.~\ref{fig:Geometrical_interpretation} represents graphically the solution of Eq.~\ref{eq:Stationary_density} with $\zeta = C_{m} k$, for two distinct values of $k$.
Theorem $1$ establishes general results for the local stability of the fixed points.

\begin{theorem}
Let $\mathcal{G}$ be a simple connected and undirected network made of $\Omega$ nodes $v_1,v_2,\cdots,v_{\Omega}$ such that their corresponding degrees are sorted in increasing order, i.e. $k_1\leq k_2\leq \cdots \leq k_\Omega$. 
	%Suppose moreover that $\forall~ 1\leq p\leq m$ the subset of nodes ${v_p, v_{p+1},\cdots ,v_{m}}$ is a connected subgraph. 
Consider the following dynamical system:
\begin{equation}
\frac{\text{d} \rho_i}{\text{d}t}=\sum_{j=1}^{\Omega}\mathcal{L}_{ij}\Big[\rho_j g(\rho_i) - \frac{k_j}{k_i}\rho_i g(\rho_j)\Big],
\label{eq:Intersection_stationary_density}
\end{equation}
with initial conditions $\rho_i(0)=\rho_i(0)$ and $\sum_{i=1}^{\Omega} \rho_i(0) = \beta \Omega${, for some $\beta \in (0,1)$}. Suppose that $g$ is a $C^1$ concave function on the open interval $(0,1)$ with $g(0)=0=g(1)$ and $g(x)>0\,\forall x\in(0,1)$. Assume moreover that all the non-empty nodes in the steady state form a single \JFadd{community}. Then:
	%Let $\rho_j^*=\phi(c,k_j)$ be the unique non-zero solution of the equation:
    %\begin{equation}
	%    \f{\rho_j^*}{ck_j}= g(\rho_j^*),
	%\end{equation}
\begin{enumerate}
	%    \item There is a constant $c>0$ solution of:
	%\begin{equation}
	%\f{\beta m}{c}= \sum_{k_i>\f{1}{cg'(0)}}k_i\,g(\phi(c,k_i)),
	%\label{eq:valuec}
	%\end{equation}
	\item The following fixed point:
\begin{equation}
\left\{
\begin{array}{ll}
\rho_i^*=0~~\forall i : k_i<\f{1}{cg'(0)} \\
%\rho_i^*=\phi(c,k_i)~~ \forall i : k_i>\f{1}{cg'(0)}
\rho_i^*>0~~ \forall i : k_i>\f{1}{cg'(0)}
\end{array}
\right.,
\label{eq:stable_state}
\end{equation}
with $\rho_i^*>0$ the non-zero solution of (\ref{eq:Stationary_density}) is a locally stable fixed point of the above dynamical system, with the constant $c$ given by:
\begin{equation}
	c=\frac{\beta\Omega}{\sum_{j=1}^{\Omega}k_j g(\rho_j^*)}\, ,
\end{equation}
{and it coincides with $C_1$ defined from~\eqref{eq:Stationary_density}.}
	%if
	%\begin{equation}
	%    g'(\phi(c,k_i))-\f{1}{ck_i}<0~\forall i : k_i>\f{1}{cg'(0)}
	%\end{equation}
	%and provided that the set of filled nodes forms a connected subgraph.
\item Any fixed point with $\rho_i^*=0$ for some $k_i>\f{1}{cg'(0)}$ is locally unstable.
	
\end{enumerate}
\end{theorem}
	
The local stability analysis will be carried out by computing the jacobian matrix $J$ associated to the previous system:
\begin{align}
J_{ij}&=\frac{\partial}{\partial \rho_j}\left(\frac{\text{d} \rho_i}{\text{d}t} \right)\nonumber\\
&=\sum_{l=1}^{\Omega}\frac{A_{il}}{k_l}\left[
\delta_{lj} g(\rho_i)
+\rho_l g'(\rho_i)\delta_{ij} -\delta_{ij}g(\rho_l) - \rho_i g'(\rho_l)\delta_{ij}
\right]\nonumber\\
& = A_{ij}\left[\f{g(\rho_i)}{k_j} - \f{\rho_i g'(\rho_j)}{k_i}\right]\nonumber + \delta_{ij} \sum_{l}A_{il}\left[ 
\f{\rho_l g'(\rho_i)}{k_l} - \f{g(\rho_l)}{k_i}
\right].
\end{align} 
Let us remember that the nodes are ranked in increasing order of their degrees, i.e. $k_1\leq k_2\leq k_{\Omega}$. Let $s\in \{1,2,\cdots, \Omega\}$ be such that: 

\renewcommand{\arraystretch}{1.5}
\begin{align}
	\left\{
	\begin{array}{ll}
	k_i c g'(0)<1 ~ & \text{for }i=1,\cdots,s \\
	k_i cg'(0)>1~& \text{for } i=s+1,\cdots,{\Omega}
	\end{array}
	\right. .
\end{align}

Let us now evaluate the jacobian matrix at the fixed point given by Eq.~\eqref{eq:stable_state}. We will denote by $M$ the jacobian matrix $J$ evaluated at this fixed point. The fixed point will be locally stable if and only if all the eigenvalues of $M$ are negative.

\begin{enumerate}
    \item For $i,j=1,\cdots,s$, we have $\rho_i^* = 0$ and $\rho_j^* = 0$ and thus:
    \begin{align}
        M_{ij} &= \delta_{ij} \sum_{l=s+1}^{\Omega} A_{il} \left[\f{\rho_l^* g'(0)}{k_l}-\f{g(\rho_l^*)}{k_i} \right]\nonumber\\
         &= \delta_{ij} \sum_{l=s+1}^{\Omega} \f{A_{il} \rho_l^*}{k_l}\left[g'(0)-\f{1}{ck_i}\right].
    \end{align}
    Since $k_i c g'(0)<1$ for $i=1\cdots,s$, we deduce that $M_{ii} < 0$ for $i=1,\cdots,s$.
    %\begin{equation}
	%M_{ii} < 0 ~\text{and}~J_{ij} = 0 ~\text{ for }~i=1,\cdots,s.
    %\end{equation}
    We also deduced that any fixed point with $\rho_i^*=0$ and $k_i c g'(0)>1$ will be unstable. 
    
    \item For $i=1,\cdots,s$ and $j=s+1,\cdots,\Omega$, we have $M_{ij} = 0$.
    Consequently $M$ is a block matrix and it remains to investigate the stability of the submatrix $(M_{ij})_{ij}$ with $i,j=s+1\cdots,\Omega$.
    
    \item For $i,j=s+1,\cdots,\Omega$, we have :
    \begin{align}
        M_{ij} &= \f{A_{ij}\rho_i^*}{k_i}\left[\f{1}{ck_j}-g'(\rho_j^*)\right]\nonumber+\\&+\delta_{ij} \sum_{l=s+1}^{\Omega}\f{A_{il}\rho_l^*}{k_l} \left[g'(\rho_i^*)-\f{1}{ck_i}\right].
    \end{align}
    Let us observe that:
    \begin{align}
        \sum_{i=s+1}^{\Omega} M_{ij} &= \sum_{i=s+1}^{\Omega} \f{A_{ij}\rho_i^*}{k_i}\left[\f{1}{ck_j}-g'(\rho_j^*)\right] \nonumber+\\&+\sum_{l=s+1}^{\Omega} \f{A_{jl}\rho_l^*}{k_l}\left[g'(\rho_j^*)-\f{1}{ck_j}\right]=0,
    \end{align}
since the network is undirected. Moreover, $\sum_{\substack{i=s+1\\ i\neq j}}^{\Omega} M_{ij} = -M_{jj}$. Assuming $g'(\rho_i^*)-\f{1}{ck_i}<0~\forall i=s+1\cdots,\Omega$, one can make use of the Gershgorin theorem to deduce the stability of the submatrix $(M_{i,j})_{ij=s+1}^{\Omega}$. Indeed, according to this theorem, the spectrum of this submatrix is contained in the set of disks
\begin{equation}
    \bigcup\limits_{j=s+1}^{\Omega}D(M_{jj},R_j)
\end{equation}
centred in $M_{jj}<0$ and with radius
\begin{equation}
    R_j = \sum_{\substack{i=s+1\\ i\neq j}}^{\Omega} \vert M_{ij}\vert  = -M_{jj}.
\end{equation}
\end{enumerate}
Consequently, we deduce that the spectrum of the submatrix $(M_{i,j})_{ij=s+1}^{\Omega}$ is contained in the left half plane, hence showing that all the eigenvalues of the matrix $M$ have a negative real part.   

\section{Detailed analysis of \JFadd{metapopulations} breaking}
\label{App_B}

\begin{figure}
	\centering
	\includegraphics[scale = 0.5, trim = 0 0 0 0]{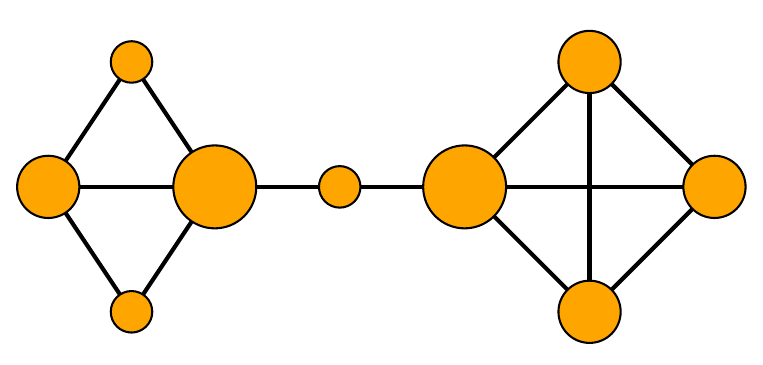}
	\caption{As the reaction-diffusion process takes place, the first node(s) to become empty is (are) the one(s) with the lowest degree, namely $k_{\mathrm{min}}=2$, for the above graph.}
	\label{fig:toy_nw}
\end{figure}

\JFadd{In this Appendix, we further investigate the role of the network topology on the emergence of (sub)communities. Let us recall that the dynamical system is given by}:
\begin{equation*}
\frac{\text{d} \rho_i}{\text{d}t}=\sum_{j=1}^\Omega\mathcal{L}_{ij}\rho_i\rho_j\left[1-\rho_i - \dfrac{k_j}{k_i}(1-\rho_j)\right]\, , \, \forall i = 1,\dots,\Omega\, ,
\end{equation*}
with $g(x)=x(1-x)$.
For sufficiently large values of the average density $\beta$ and generic initial conditions without empty nodes, there will be no asymptotically empty nodes, in which case the stationary nodes densities are given by
\begin{equation*}
\rho_i^* = 1-\frac{1-\beta}{k_i \langle 1/k \rangle},
\end{equation*}
where $\langle 1/k \rangle = \sum_{j=1}^{\Omega} \frac{1}{k_j}$. Empty nodes emerge as soon as there is {at least} a node of degree $k_i$ such that $k_i{=}\frac{1-\beta}{\langle 1/k \rangle}$. Hence, the first node(s) to become empty are the ones with the lowest degree $k_{\mathrm{min}}$ and this happens when $\beta = 1-k_{\mathrm{min}} \langle 1/k \rangle$.  As an example let us consider the network model of Fig. 1 in the main text that we here repropose as Fig.~\ref{fig:toy_nw} for which $k_{\mathrm{min}}=2$. There are three nodes with degree $2$ whose density could potentially vanish for some $\hat{\beta}$. A simple computation returns $\hat{\beta}=1-\frac{20}{27}\sim 0.259$.\newline

%Let us consider again the network structure given in \ref{fig:metacommunities}. It is possible to predict the value of the average \JFdel{node density} \JFadd{population density} $\beta$ for which the first empty node will emerge. The}

\begin{figure*}
	\centering
	\includegraphics[width=\textwidth]{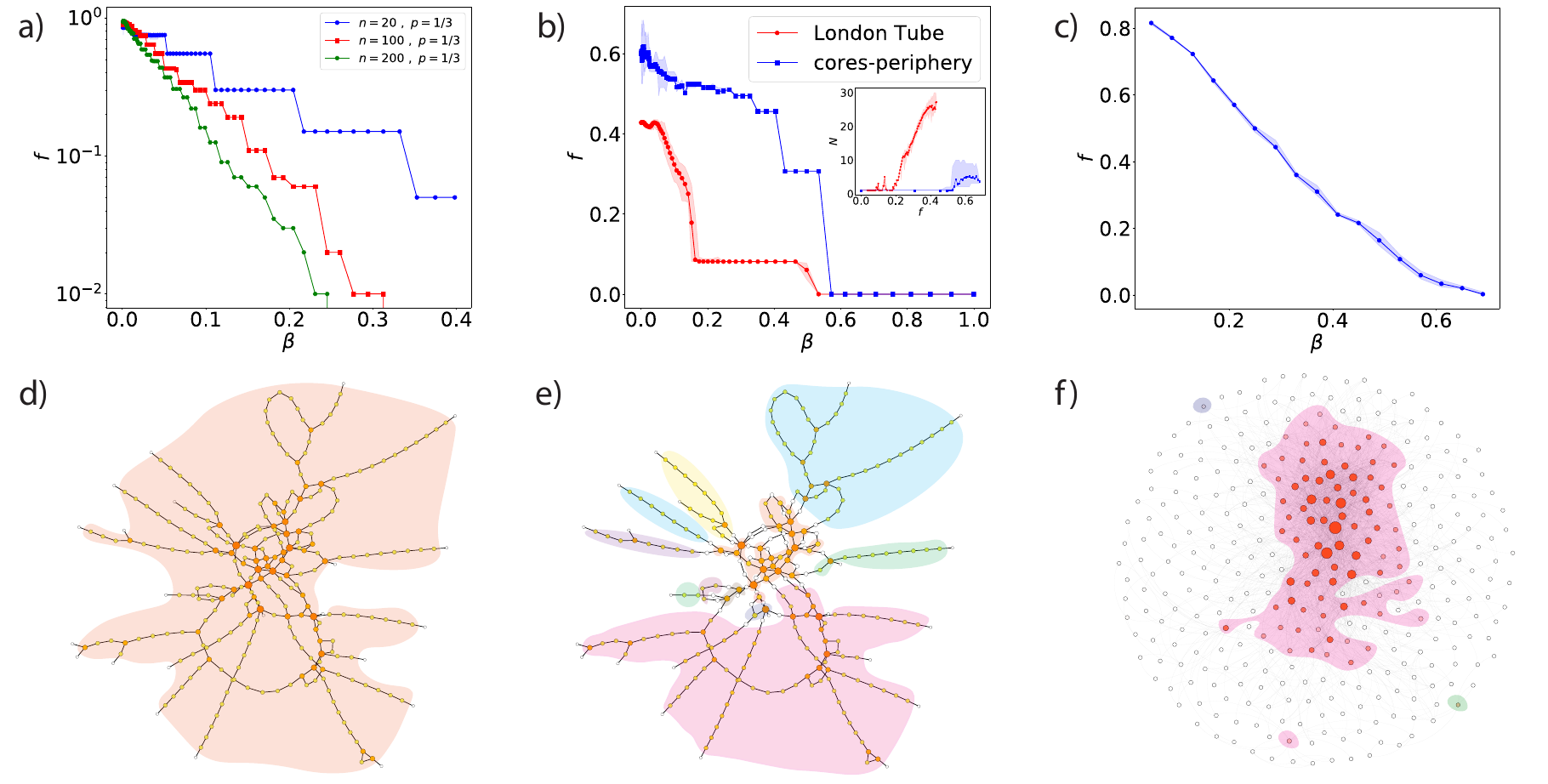}
	\caption{ (Upper panels) Fraction of empty nodes $f$ vs. the average density $\beta$. The shaded {blue and red} areas (in panels b) and c)) show the multistability {phenomenon} as measured by the min-max {node} occupancy resulting from {several independent simulations}. $\textbf{a)}$ Erdos-Renyi networks with different network sizes and same probability to have a link among two nodes ($p=1/3$). As the number of nodes increases, the quantisation becomes weaker while the multistability is almost absent. $\textbf{b)}$ The London Tube network \cite{de_domenico_navigability_2014} (red curve) vs. a synthetic core-periphery one (blue curve). The shaded area corroborates the multistability claim, and the quantisation appears neatly {manifested by the plateaus}. (Inset) Number of \JFadd{(sub)communities} $N$ vs. the fraction of empty nodes $f$ showing that to the same fraction of empty nodes corresponds a different number of \JFadd{(sub)communities}. $\textbf{c)}$ The disassortative USAir97 network \cite{nr}. (Lower panels) \JFadd{(Sub)communities} visualisation (shaded coloured areas) and empty nodes (white circles): $\textbf{d)}$ and $\textbf{e)}$ results for two different densities initialisations for the same value of $\beta=0.15$ close to the abrupt transition in $f$ vs $\beta$ (see panel $\textbf{b)}$) in the London Tube network. $\textbf{f)}$ the same result of panels $\textbf{d)}$ and $\textbf{e)}$ for the USAir97 network for $\beta=0.15$.}
	\label{fig:nested}
\end{figure*}

\JFadd{To better infer the role of network topology on both the \JFadd{metapopulation} breaking and the quantisation of the fraction of vacancies, in Fig.~\ref{fig:nested} we analyse a set of both synthetic and real networks with different structural characteristics. 
%In particular, we consider Erd\H{o}s-Rényi graphs and human transport networks. 
%well-known as a medium for spreading pathogens, such as bacteria and viruses, thus enabling the dispersal between their respective \JFdel{metacommunities} \JFadd{metapopulations} \cite{tatem_global_2006, crowl_spread_2008}.
In panel $a)$, we consider Erd\H{o}s-Rényi graphs and show that as the size of the random networks increases, the quantisation effect in occupying, respectively depleting the nodes becomes more diluted.
The importance of the heterogeneity of the degree distribution in the formation of (sub)communities is further accentuated in diassortative networks where nodes are naturally arranged in interacting groups of highly and loosely connected entities. In panel $b)$, we consider a synthetic and a real core-periphery network, where at variance with the Erd\H{o}s-Rényi topology (of panel $a)$), besides quantisation, we observe the emergence of many different (sub)communities (panels $d)$ and $e)$). This scenario is a neat manifestation of multistability, i.e., different stable states for different initial conditions for the same {system parameter $\beta$}, as attested by the shaded area in panel $b)$. In fact, for two initial nodes densities {conditions}, with same total mass, randomly chosen from a uniform distribution, we have two distinct configurations of (sub)communities with a different fraction of vacant nodes. There is a clear difference between the synthetic and the empirical network since the latter is provided with ``bridge'' nodes that connect two or more cores. Let us recall that the bridgeness centrality of node $v_i$ is the fraction of shortest paths starting and ending not in the neighbourhood of node $v_i$ and running through it \cite{jensen2016detecting}. This peculiarity makes the real core-periphery network prone to split in different (sub)communities once the density $\beta$ starts to decrease, at variance with the synthetic network where multiple (sub)communities appear only when the density takes low values.
In panels $c)$ and $f)$, we consider the USAir97 network, an undirected network whose links represent the US air flights in $1997$ \cite{nr}. The network was shown to be a core-periphery network with a single densely connected core \cite{fasino2020fast}. As the parameter $\beta$ decreases, most of the mass concentrates in the core of the network, as shown in panel $f)$, leaving a single large community of occupied nodes. Such difference between the behaviors arising from the two networks can be justified by the presence of bridge nodes that allow a broader repartition of the mass $\beta_m$.}

\begin{figure}[htb!]
		\centering
		\includegraphics[scale = 0.4, trim = 0 350 0 0]{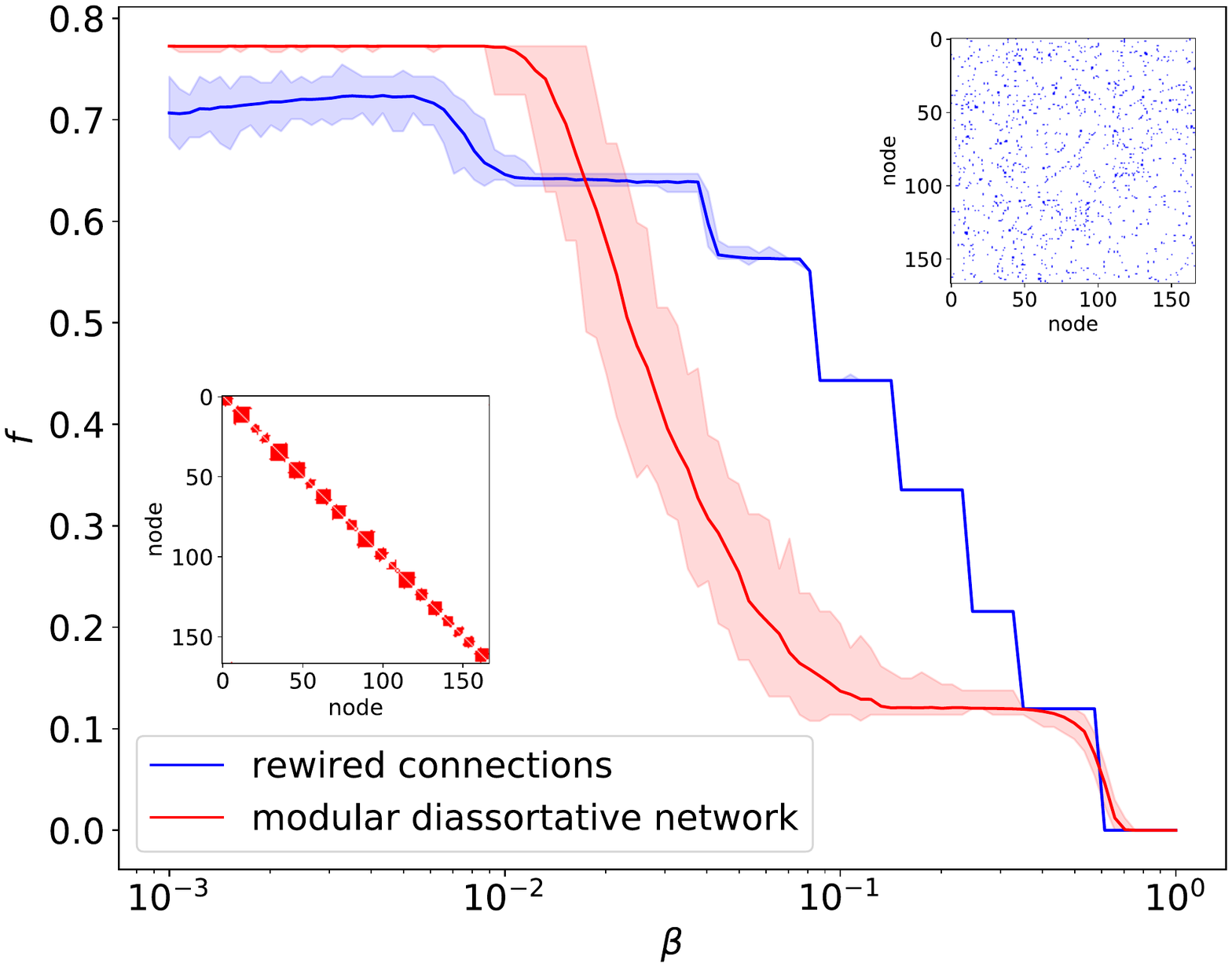}
		\caption{Fraction of empty nodes $f$ vs. the average density $\beta$. The results were averaged over $100$ distinct {network realisations}. The shaded area shows the multistability as measured by the min-max occupancy resulting from all the realisations.}
		\label{fig:f_modular}
\end{figure}

To further elucidate this assertion, in Fig.~\ref{fig:f_modular}, we analyse the impact of the {network modularity} on the fraction of empty nodes $f$. The red curve shows $f$ vs. $\beta$ for a disassortative modular network made of $20$ densely connected subgraphs {formed by a number of nodes comprised between $5$ and $10$}, linked together through bridge nodes, i.e. nodes {whose removal splits the network into disconnected subgraphs.} Such networks can be obtained starting from a {ring} network and then replacing every two nodes by a densely connected subgraph. The corresponding adjacency matrix is shown in the inset on the lower left corner and confirms the modularity of the network (each pixel represents a connection between $2$ nodes). As the average density $\beta$ decreases, we observe that the mean fraction of empty nodes increases in a continuous way {up to reaching a first plateau corresponding to the case where (almost all the) bridge nodes are empty and the network is split into (almost) $20$ \JFadd{subcommunities}. For smaller values of $\beta$ the fraction of empty nodes increases again} until it stabilises for low values of $\beta$, {i.e. smaller than $\sim 10^{-2}$, when also nodes into the subgraphs are empty}. The blue curve shows $f$ vs. $\beta$ for a network obtained using a rewiring degree preserving process on the former one. The new adjacency matrix is now represented in the inset on the upper right corner. Due to the rewiring procedure, the network now looks like a random network but the degree sequence is unchanged. The corresponding behaviour of $f$ vs. $\beta$ is now quite different and exhibits a step-like behaviour. In both cases, the results were averaged over $100$ distinct {network realisations}. The min-max deviation {in the resulting value of $f$} is represented by the dashed region.\newline

%we have introduced in the following a simple model of core-periphery network with a tunable fraction of bridge nodes, obtaining qualitatively identical results with the London Tube network \cite{de_domenico_navigability_2014}.\\}

%(for $100$ logarithmically equally spaced values of $\beta$). 
	
\begin{figure}[htb!]
	\centering
	\includegraphics[scale = 0.35, trim = 0 25 0 0]{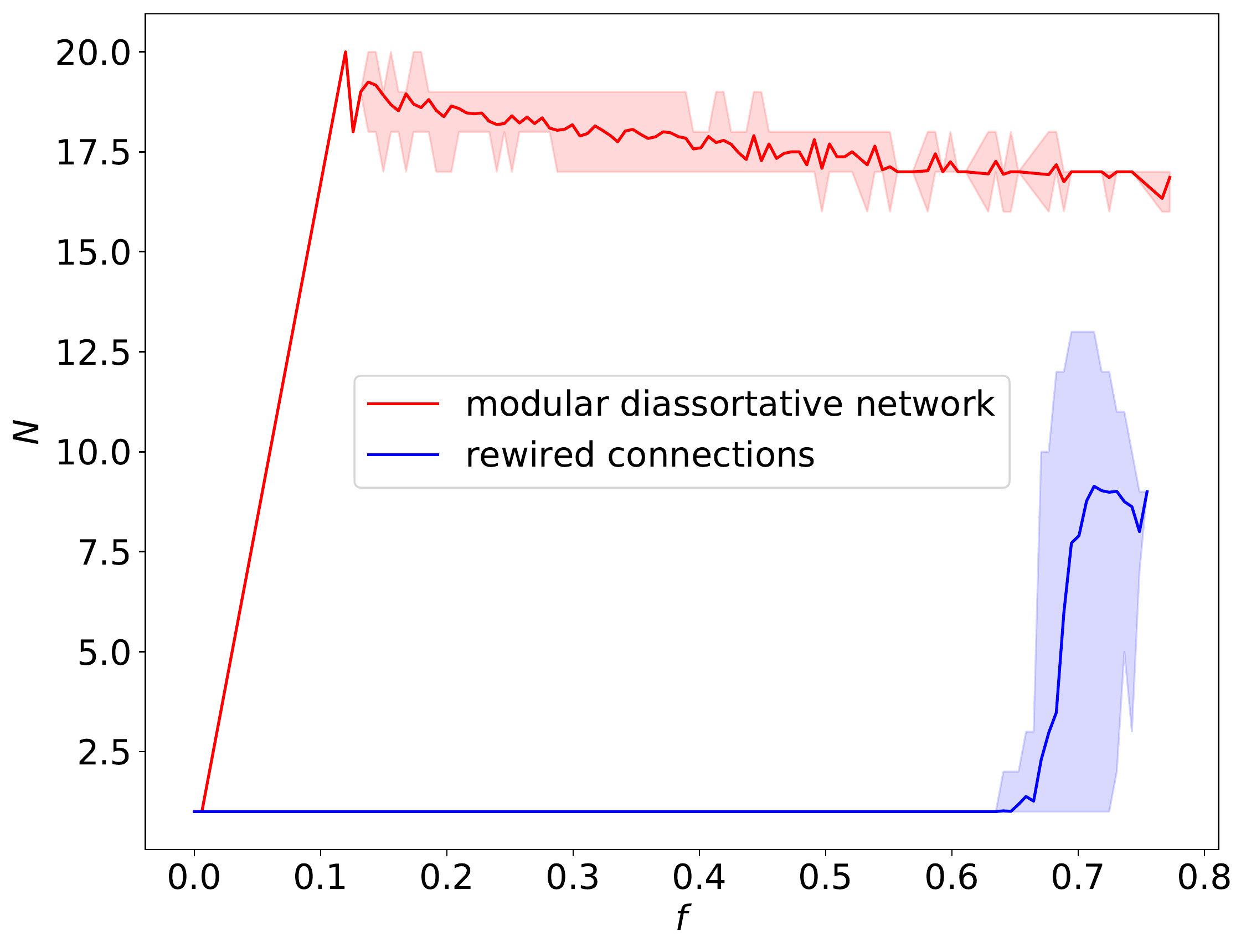}
	\caption{Number $N$ of \JFadd{(sub)communities} vs. the fraction of empty nodes $f$. The shaded area shows the multistability as measured by the min-max deviation resulting from all the realisations.}
	\label{fig:Mass}
\end{figure}

\begin{figure}[htb!]
	\centering
	\includegraphics[scale = 0.4, trim = 0 25 0 0]{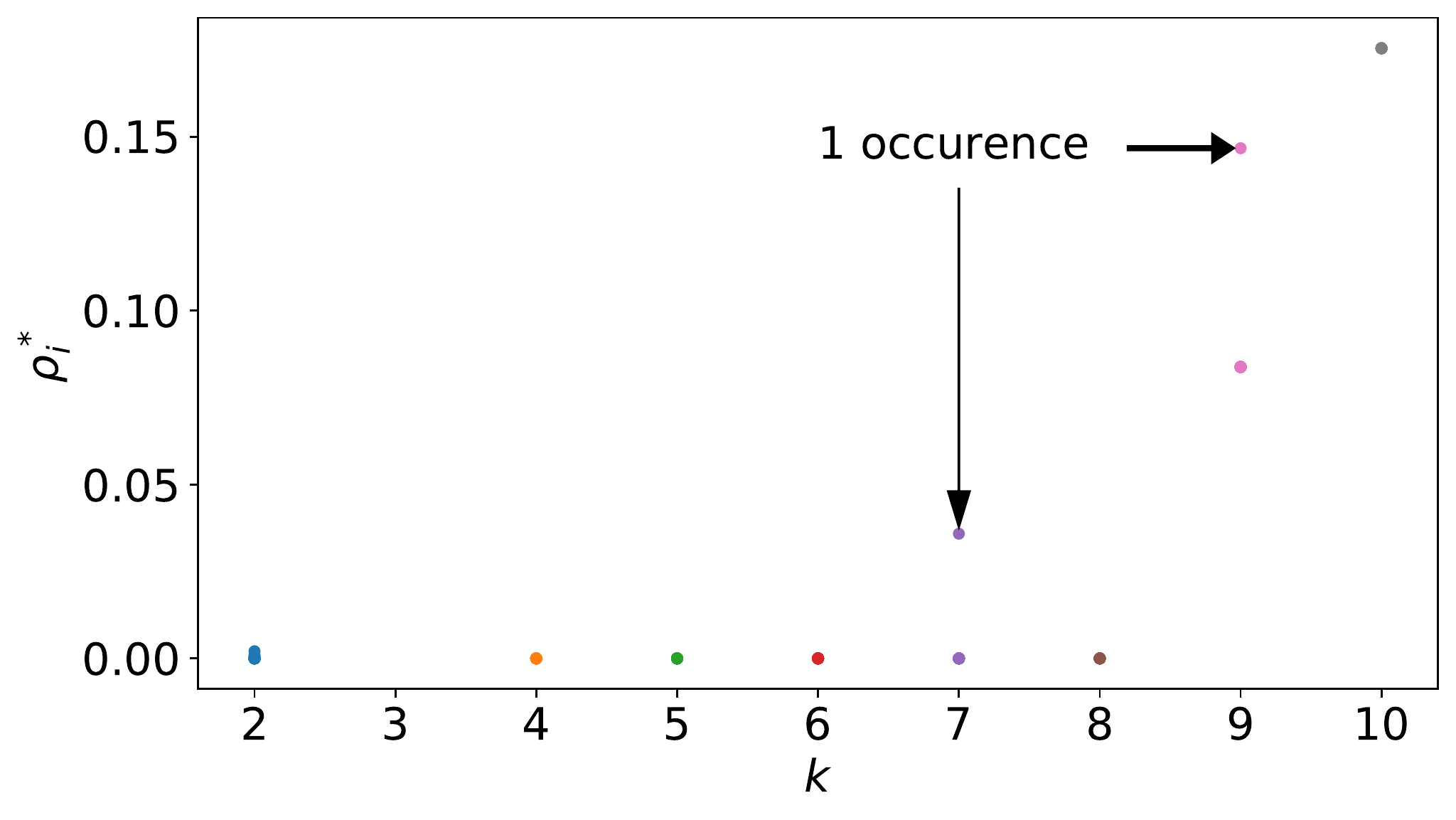}
	\caption{Stationary nodes densities $\rho_i^*$ as a function of their degree $k$, for the rewired version of the modular disassortative network used in Fig.~\ref{fig:f_modular}. The initial conditions are given by $\rho_i(0) = 0.15$. {Nodes densities were initialized randomly such that the average density satisfies $\beta = 0.15$.}}
	\label{fig:Mass_r}
\end{figure}

\begin{figure}[htb!]
	\centering
	\includegraphics[scale = 0.4, trim = 0 25 0 0]{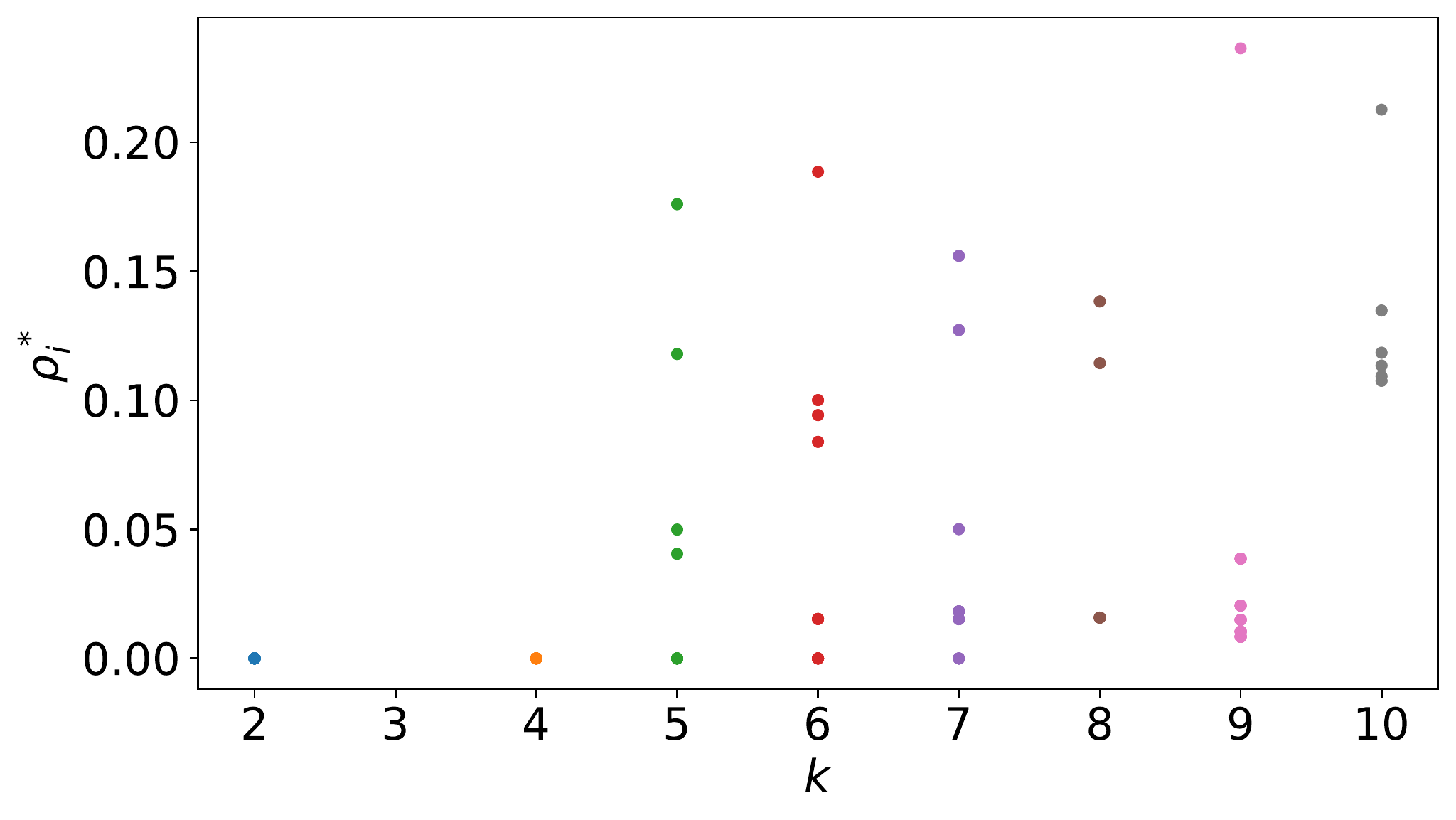}
	\caption{Stationary nodes densities $\rho_i^*$ as a function of their degree $k$, for the modular disassortative network used in Fig.~\ref{fig:f_modular}. {Nodes densities were initialized randomly such that the average density satisfies $\beta = 0.15$.}}
	\label{fig:Mass_2}
\end{figure}

{To understand these different behaviours we analyse the number of \JFadd{(sub)communities} emerging in the modular network as well as in the rewired one (see Fig.~\ref{fig:Mass})}. {{For a very large range of values of $\beta$} the rewired network {exhibits} one \JFadd{(sub)community} (see the blue curve in Fig.~\ref{fig:Mass}), associated to a unique constant $C_1\equiv C_1(\beta)$. Consequently, nodes densities are essentially constrained by their degree, as shown in Fig.~\ref{fig:Mass_r} where we reported the stationary nodes densities $\rho_i^*$ vs. their degree (nodes densities were initialized randomly such that the average density satisfies $\beta = 0.15$). As a consequence, nodes with a degree larger than $C_1$ are filled while all the others become empty. Since the degree sequence is finite, this implies the observed jumps on the blue curve given in Fig. \ref{fig:f_modular}. On the other hand, for the diassortative modular network, the mass first distributes among the distinct modules ($20$ in this case) leaving the bridges nodes empty (see the red curve in Fig. \ref{fig:Mass}). To each of these modules corresponds a constant $C_m=\left(1-\beta_m\right)/\langle 1/ k\rangle_{\Omega_m}$. Inside the $m$-th module, all the nodes with a degree $k_i<C_{m}$ will then become empty. As $C_{m}$ is of order $C_1/l$, with $l$ the number of modules, nodes with smaller degree can also be filled, leading to a lower value of the fraction of empty nodes, except when $\beta$ is very large or very small (see Fig. \ref{fig:f_modular}). Moreover, as the constants $C_{m}$ differ from each other, distinct stationary densities can be observed for nodes with the same degree (see Fig. \ref{fig:Mass_2}). Since the way the mass distributes among the modules depends (in a highly non-trivial way) on the initial densities, averaging over sufficiently many configurations leads in fine to a smooth evolution of $f$ vs. $\beta$ (see the red curve in Fig. \ref{fig:f_modular}).}\newline 
	
\begin{figure}[htb!]
	\centering
	\includegraphics[scale = 0.5, trim = 0 0 0 0]{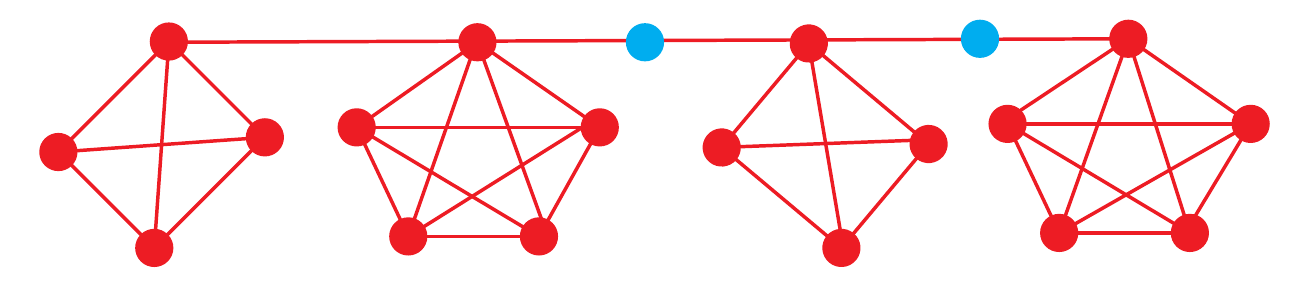}
	\caption{Modular network with cliques of size $k=4,5$ connected through paths of length $1$ or $2$. The removal of a (blue) node belonging to a path of length $2$ breaks the connectivity of the network.  }
	\label{fig:chain_network}
\end{figure}

\begin{figure}[htb!]
	\centering
	\includegraphics[scale = 0.35, trim = 0 25 0 0]{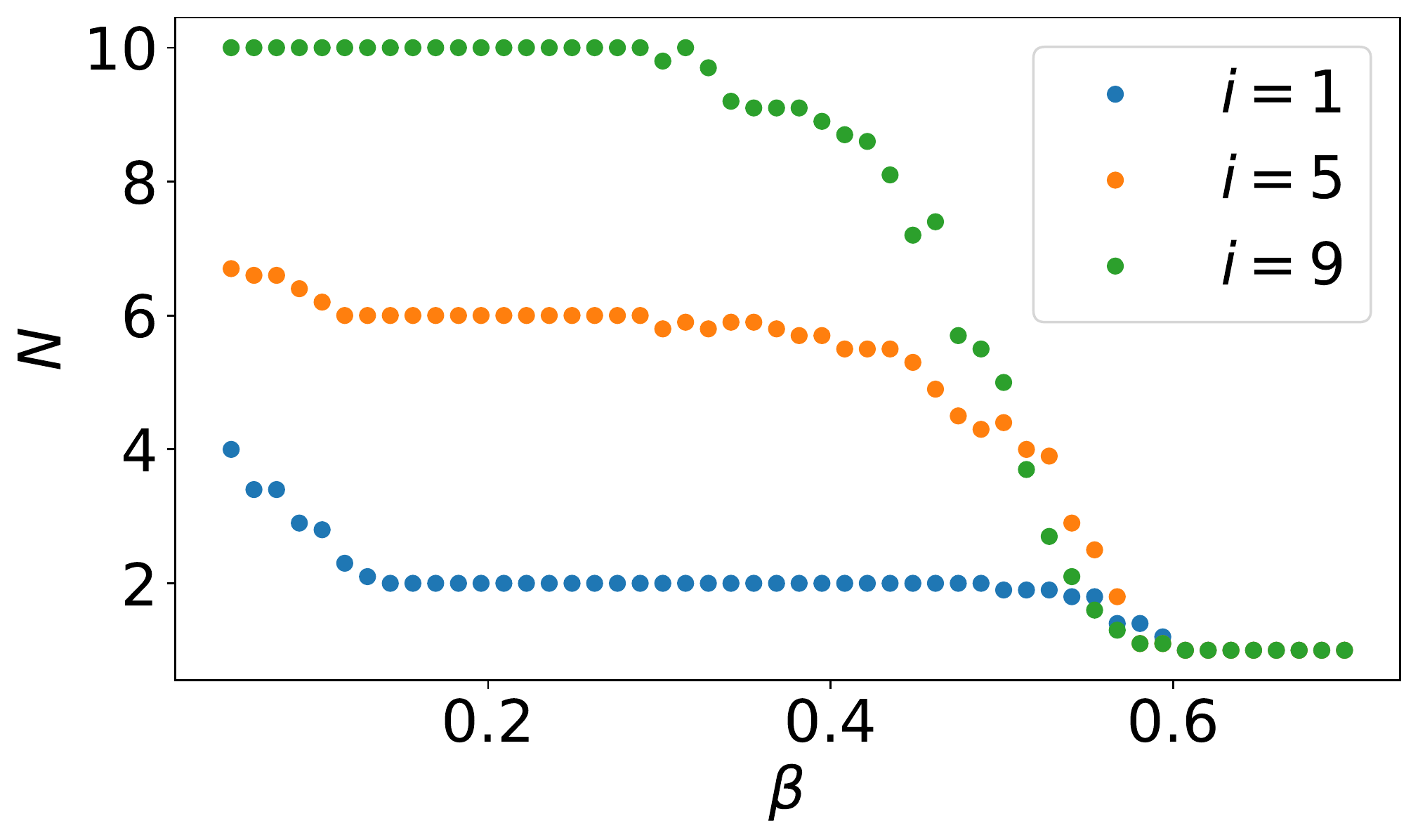}
	\caption{Number of \JFadd{(sub)communities} $N$ as a function of the average node density $\beta$, for a modular network with a number $i$ of bridge nodes. The network was built using the procedure described in the text with $10$ cliques of size $4\leq k \leq 7$. For each value $i$, the data were averaged over $10$ independent initial configurations.}
	\label{fig:chain}
\end{figure}

%\newpage
To highlight the impact of node bridgeness in the emergence of \JFadd{(sub)communities}, we show in Fig. \ref{fig:chain} the results obtained using a small modular network with a varying fraction of bridge nodes of degree $2$ between the modules and we computed the number of \JFadd{(sub)communities} $N$ vs. $\beta$. To construct such network, we first generate $m$ cliques of size $k$ ($k\geq 3$). We then connect the $i$-th clique to the $(i+1)$-th clique ($i=1,\cdots,m-1$) by means of a path of length $1$ or $2$. The procedure is illustrated in Fig. \ref{fig:chain_network}, for $m=4$ and $k=4,5$. As $\beta$ is reduced, the bridge nodes become empty, leaving the mass in the disconnected modules. The larger the number of bridge nodes, the larger the number of \JFadd{(sub)communities}.\newline

\begin{figure}[htb!]
	\centering
	\includegraphics[scale = 0.25]{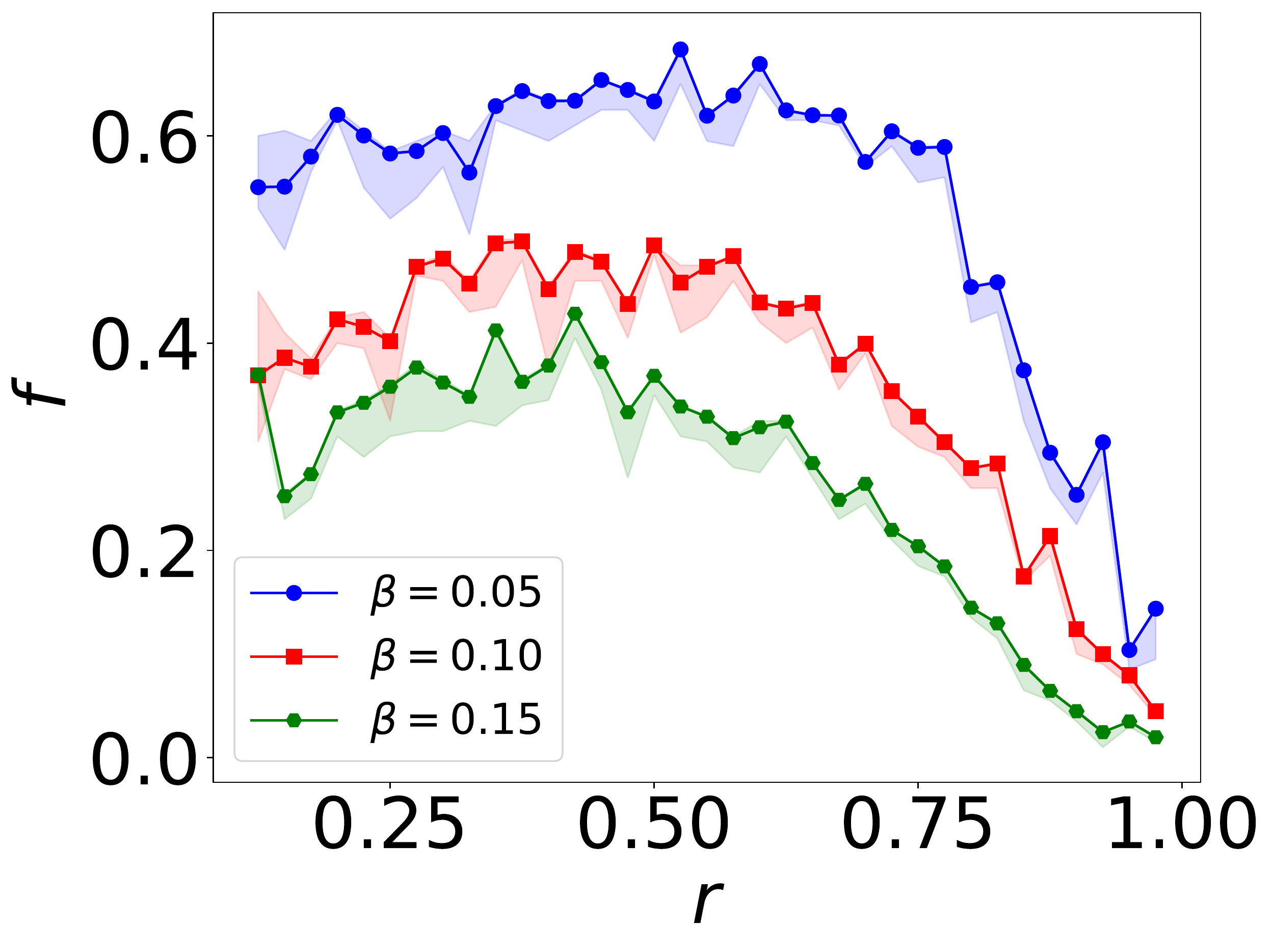}
	\caption{\JFadd{Fraction of empty nodes in random geometric graphs obtained by connecting every pair of (uniformly distributed) points whose euclidean distance is smaller than a given thresold $r$. The larger $r$ and the smaller $f$.}}
	\label{fig:RGG_radius}
\end{figure}

\JFadd{We end up this Appendix by considering random geometric graphs obtained by starting with a set of uniformly distributed points in the unit square and connecting them if their euclidean distance is less than a given thresold $r$. The larger $r$ and the smaller the fraction of empty nodes as shown in Fig. \ref{fig:RGG_radius}. In particular, for $r=\sqrt{2}$, we obtain a complete graph, for which there is no empty node.}

\clearpage

\bibliographystyle{apsrev4-2}%{abbrv}%{unsrtnat}%{plainnat}%{abbrvnat}%{apsrev4-2}
\bibliography{biblio}% Produces the bibliography via BibTeX.

\end{document}